# Physics mechanisms of fines detachment and migration during CO2-water corefloods

C. Nguyen[1], G. Loi[1], T. Russell[1], Y. Yang[1], N.N. Zulkifli[2], M. I. Mahamad Amir [2], A.A. Abdul Manap[2], S.R. Mohd Shafian[2], A. Badalyan[1], P. Bedrikovetsky[1], and A. Zeinijahromi[1]

[1]School of Chemical Engineering, The University of Adelaide, Adelaide, South Australia, Australia

[2]Petroliam Nasional Berhad, Kuala Lumpur, Malaysia

**Key points:**

- Eight corefloods on displacement of water by CO2 exhibit unstable gas viscous fingering.
- Permeability decline is attributed to fines migration and some salt precipitation.
- Fines detach by moving CO2-water menisci; pendular water rings secure the fines.
- Water is produced during $10^3$ PVIs, full evaporation occurs during $10^6$ PVIs.
- Abrupt gas permeability increase is explained by reaching the percolation threshold by gas saturation.

**Abstract**

One of the key risks for a Carbon Capture Storage (CCS) is injectivity decline. Evaporation of the connate brine in near-wellbore region during $CO_2$ injection may result in drying-up the rock yielding the mobilisation and migration of clay particles leading to decline rock permeability and consequent loss of well injectivity. Influx of the reservoir brine into the dried-up zone yields accumulation of precipitated salt and injectivity decline. This paper presents the results of eight coreflooding experiments aiming investigation of the effect of rock dry-out, fines migration, and salt precipitation during $CO_2$ injection. Pressure drops across the cores, brine saturation and produced clay fines concentration versus Pore Volume Injected (PVI) have been measured.

All lab tests exhibit the following features: intensive fines production at the very beginning of gas-water production period following reduced-rate fines production during overall evaporation period and continuous fines disappearance at the late stage; abrupt increase in gas permeability in the middle of evaporation, and non-monotonic evaporation rate and pressure drop. To explain these phenomena, we distinguished three sequential regimes of fines detachment during two-phase displacement: (i) moving gas-water menisci; (ii) pendular rings of residual water; (iii) dry flux, and found that for the conditions of our corefloods, detachment is possible in regime (i) only. Fines production during overall evaporation period is explained by simultaneous occurrence of three regimes during unstable displacement of water by gas in micro-heterogeneous rock.

**Nomenclature**

**English letters**

| | |
|---|---|
| $F_c$ | Capillary force, ML/T$^2$ |
| $F_d$ | Drag force, ML/T$^2$ |
| $F_e$ | Electrostatic force, ML/T$^2$ |
| $F_g$ | Gravitational force, ML/T$^2$ |
| $F_l$ | Lift force, ML/T$^2$ |
| $h$ | Particle-substrate separation distance, L |
| $k_{rg(s)}$ | Gas relative permeability |
| $p$ | Pressure, ML$^{-1}$T$^{-2}$ |
| $r_s$ | Particle size, L |
| $S_w$ | Water saturation |
| $U$ | Velocity, LT$^{-1}$ |
| $V$ | Energy, ML$^2$T$^{-2}$ |
| $V_r$ | Volume of pendular ring, L$^3$ |

**Greek letters**

| | |
|---|---|
| $\Delta$ | Difference |
| $\mu_g$ | Gas viscosity, ML$^{-1}$T$^{-1}$ |
| $\mu_w$ | Water viscosity, ML$^{-1}$T$^{-1}$ |

**Abbreviations**

| | |
|---|---|
| BTC | Breakthrough concentration |
| CO$_2$ | Carbon dioxide |
| DLVO | Derjaguin-Landau-Verway-Overbeek Theory |
| HPLC | High performance liquid chromatography |
| PDF | Probabilistic distribution function |

## 1. Introduction

Carbon capture and geological storage (CCS) is a key pillar in efforts to put the world on the path to net-zero emissions and is one of the few methods that can remove CO$_2$ from the atmosphere at an industrial scale. One of the fundamental problems while injecting and storing large volumes of carbon dioxide in subsurface formations is well impairment due to decrease in permeability and consequent injectivity decline; this can adversely impact a CO$_2$ storage project and result in extensive remediation costs and project failure.

Formation damage (permeability decline) and injectivity impairment during CO$_2$ storage in aquifers and depleted gas and oil fields has been widely reported based on field pilots and lab studies. One of the main physics mechanisms of permeability damage during CO$_2$ injection is fines migration [1, 2]. Natural reservoir fines attached to the rock surface are mobilised by either capillary force or drag, migrate in the porous space until being captured in thin pore throat, resulting in plugging the flow paths and significant permeability decline [3-6].

Fig. 1 shows the forces exerting the particle attached to the rock surface: drag, lift, capillary, electrostatic, and gravitational forces. The main detaching force during drainage with two-phase gas-brine flow is the capillary force exerting the interception curve of water-gas meniscus and the particle (Fig. 1a). However, the capillary force exerting the particle from the pendular ring of the residual water attracts the particle to the substrate (Fig. 1b). After complete water evaporation into injected $CO_2$, the particle is detached by the drag (Fig. 1c).

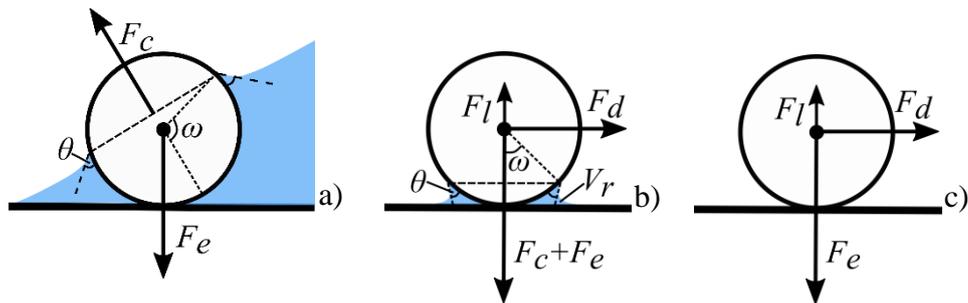

Fig. 1. Schematic for fines detachment and water evaporation during brine displacement by $CO_2$: a) capillary force exerts the particle from meniscus passing by the particle; b) particle attachment by residual brine pellicular; c) dry flow in large pores with complete fines detachment by drag.

Numerous experimental works study two-phase $CO_2$-brine transport in porous media. The brief incomplete list includes wettability alteration effects [7, 8], residual $CO_2$ entrapment [9, 10], chemical reactions [11], capillary hysteresis [12], changes in pore-space geometry [13, 14]. Effects of fines migration during $CO_2$ storage have been studied experimentally with respect to well injectivity [15, 16], relative permeability [17], $CO_2$ residual trapping [18], and formation damage [19]. Despite of those investigative efforts, the role of fines migration during two-phase carbon dioxide transport in aqueous environment is poorly understood. Lab studies on brine- $CO_2$ flow with fines migration control are not available. This work analyses the fines detachment and migration during laboratory $CO_2$ flooding and explains the observed phenomena.

Along with fines migration, the permeability decrease is affected by multiple physics mechanisms: unstable two-phase displacement and viscous fingering, partial miscibility of water and carbon dioxide, rock drying, and salt precipitation. A lab study exhibiting and analysing all these effects in one $CO_2$ coreflood is not available.

The present study fills the gap. We conducted eight $CO_2$ floods using five sandstones controlling fines production along with pressure drop and water saturation until complete core drying. The tests exhibit common features of non-monotonic evaporation rate and pressure drop variations, three periods of fines production with declining rates, and abrupt permeability decrease. Mechanical equilibrium equations of the attached particles including

DLVO calculations establish the particle detachment conditions for the parameter values corresponding to these tests. The observations have been explained by the interacting phenomena of unstable displacement of brine by gas, fines detachment, migration and pore straining, capillary phenomena, and salt precipitation.

The structure of the text is as follows. Section 2 presents the basics for torques and forces exerting the attached particles, including DLVO, and their calculations for the test conditions. Section 3 presents rock and fluid properties, the preparation of sandstone core plugs, results of rock mineralogical analyses, description of experimental setup, experimental procedures, data collection methods, and the test results. Section 4 provides the detailed analysis and explanation of the experimental results. Section 5 discusses the necessary modelling to solidify the interpretation of the observed phenomena. Section 6 concludes the paper.

   2.  **Brief physics introduction in two-phase suspension-colloidal transport in porous media**

During water displacement by a low-viscosity gas, gas invades water-saturated core as a set of "parallel" fingers and the water saturation passed by gas highly exceeds the usual values of connate water saturation. Therefore, the water-filled pores in the unswept zones, the pores where the gas-water menisci pass (Fig. 1a), the pores with residual water situated around the particle-surface junction (Fig. 1b), and dry pores (Fig. 1c) are present in the core simultaneously. For injected $CO_2$, water evaporates in gas. Here the attached particle is subject to capillary, electrostatic, drag, lift and gravitational forces [20-23]. The expressions for these forces and the corresponding level arms are given in section S1.

DLVO energy profiles and shown in Fig. 2a for silica rock surface and spherical particle under brine, deionised water (DI), and gas environments (red, blue, and yellow curves, respectively). Zoom in Fig. 2b shows secondary energy minimum. The electrostatic force attaches fines to the substrate in saline brine and repulses them in DI water. To predict particle detachment from the rock by either drag or capillary force, maximum value of electrostatic attraction is accounted for, which corresponds to maximum slope of energy profiles in Fig. 2.

Figs. 1a and 3a correspond to an attached particle passed by the advancing gas-water meniscus during drainage. Here $\theta$ is the contact angle for the particle, and $\omega$ is the filling angle that determines the position of meniscus-particle intersection curve. The pore is significantly larger than the particle, so the meniscus far away from the particle is considered to be flat; the meniscus is curvilinear only near to the particle to approach to its surface with the fixed contact angle. The water filling angle $\omega$ is shown in Fig. 1a. This angle is equal to $\pi$ where the meniscus touches the particle for the first time, is equal π/2 where the meniscus plane crosses the particle centre and is equal zero at the last touch of the particle by meniscus.

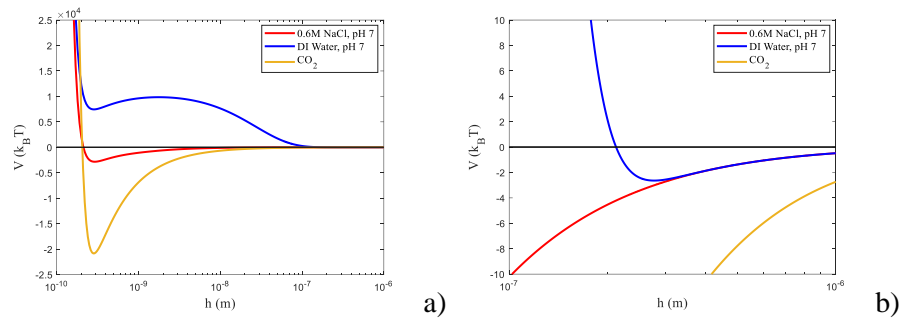

Fig. 2. DLVO energy profile for silica substrate and particles under freshwater, 0.6M NaCl brine, and gaseous $CO_2$ environments.

The blue curve in Fig. 3a that corresponds to a hydrophilic particle corresponds to particle-surface attraction for all meniscus positions; the purple curve for hydrophobic particles indicates repulsion for all meniscus positions. Here the capillary force highly exceeds the electrostatic attraction for the bulk of meniscus positions. The contact angle for carbon dioxide and silica is higher than that for inert gases or air. Yet, sandstones are water-wet in $CO_2$ environment [24, 25]. For partially-wet particles, the capillary force is repulsive at small filling angles and attractive for the large angles. Even for slightly or highly hydrophobic particles, maximum attaching and detaching capillary forces exceed the electrostatic force by one order of magnitude. The above explains the fines detachment by the capillary force exerting particle by the meniscus passing over the particle.

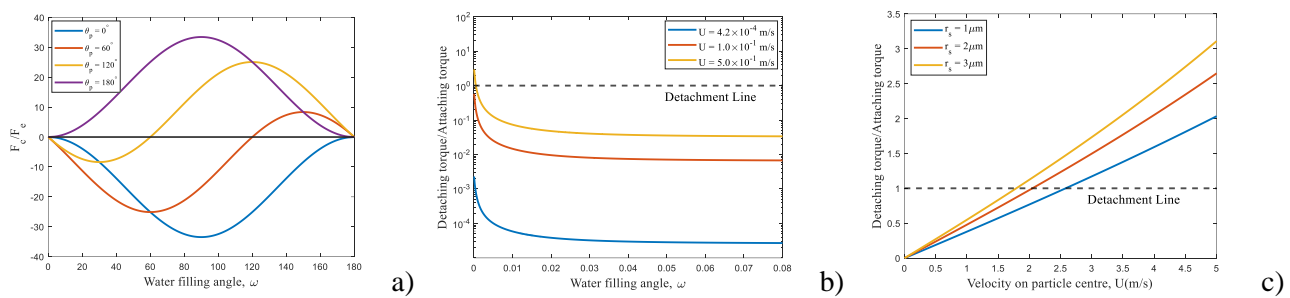

Fig. 3. Particle attachment and detachment: a) the ratio between the capillary and electrostatic forces for meniscus passing by the particle in Fig. 1a; b) the ratio between the torques of drag and the total of electrostatic and capillary forces for the brine pendular ring in Fig. 1b; c) the ratio between the drag and electrostatic torques for dry flow in Fig. 1c.

Fig. 3b shows the ratio between the detaching drag and lift torque and the attaching torque by capillary, electrostatic and gravitational forces [26]. This corresponds to Fig. 1b of the water bridge between the particle and substrate. The torque ratio is almost independent of the meniscus position and highly increase at small angles, where the meniscus almost passes the particle. Even for high flow velocity 4.2×10⁻⁴ m/s (blue curve), the attaching torque significantly exceeds the detaching torque for all menisci positions. The ratio reaches the value 0.45 for velocity U=0.1 m/s, and still the particles remain attached (red curve). Increase of velocity up to 0.5 m/s yields the particle detachment at low water filling angles (yellow curve).

For fines detachment in dry cores, Figs. 1c and 3c show the detachment by drag and lift. The ratio between the detaching torque by drag and lift, and the attaching torque by electrostatic force is presented in Fig. 3c for velocity U=4.2×10$^{-4}$ m/s. The larger is the particle the lower is the detachment velocity. For particle sizes 1, 2, and 3 μm, the detaching velocities are 2.6, 1.9, and 1.7 m/s.

### 3. Laboratory methodology

This section presents the methodology of the experimental study, including properties of rocks and fluids (section 3.1), description of the experimental setup (section 3.2), detailed experimental procedures (section 3.3), and the results of experiments (section 3.4).

**3.1. Rocks and fluids**

Five sandstone cores have been used in the tests: two low-permeable Berea cores, two medium-permeable Buff Berea cores, and one high-permeable Bentheimer core. Core dimensions, imbibition porosities, and undamaged brine and gas permeabilities are presented in Table S1.

Table S2 presents mineral composition of these cores determined by a quantitative X-ray diffraction (XRD) analysis (Fig. S1) using a Bruker D8 ADVANCE Powder X-ray Diffractometer with a Cu-radiation sources. Data were processed using Bruker DIFFRAC.EVA software and Crystallography Open Database reference patterns for identifying mineral phases. Quantification was carried out using TOPAZ profile fitting based software. Table S2 shows that the sandstone cores have low-to-moderate clay concentration; the most abundant mineral is quartz. Prior to the tests, the cores were dried in an atmospheric oven at 60°C for 24 hours, and then evacuated in desiccator under vacuum for another 24 hours until core mass stabilisation, and the initial core mass was measured. After that, the cores were saturated with a 0.6 M sodium chloride (NaCl) solution, and core imbibition porosities were measured. All solutions prepared for these tests used analytical grade NaCl and degassed MilliQ deionised water. The solutions were filtered through 0.2 μm Nuclepore Track-Etched Polycarbonate Membrane filter.

**3.2. Experimental set up**

Fig. 4 shows the schematic of the experimental set-up; Fig. S2 shows its photo. The detailed description of all elements is given in S3.

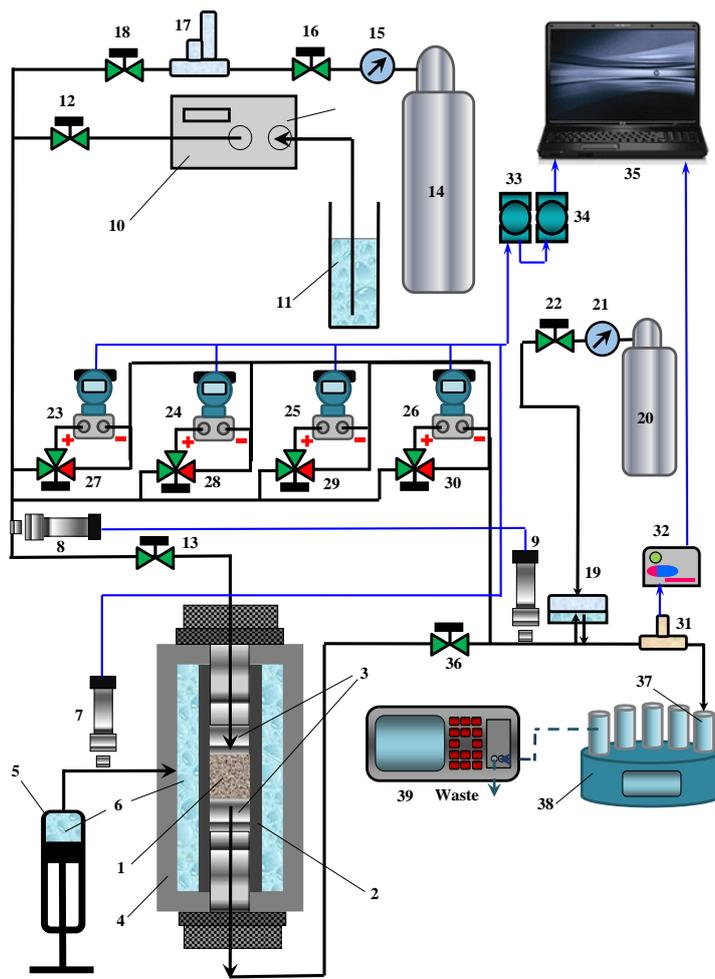

Fig. 4. Schematic of the experimental setup: 1 – Sandstone core; 2 – Elastic Viton sleeve; 3 – Flow distributors; 4 – High pressure coreholder; 5 – Overburden pressure generator; 6 – Distilled water; 7-9 – Pressure transmitters; 10 – HPLC pump; 11 – Brine solution; 12, 13, 16, 18, 22, 36 – Two-way valves; 14 – Cylinder with compressed $CO_2$ gas; 15, 21 – Gas pressure regulator; 17 – Mass flow controller; 19 – Back pressure regulator; 20 – Cylinder with compressed air; 23-26 – Differential pressure transmitter; 27-30 – Three-way valve; 31 – Electrolytic conductivity sensor; 32 – Signal transmitter; 33-34 – Data acquisition module; 35 – PC-based data acquisition system; 37 – Sampling tubes; 38 – Sampling carousel; 39 – Particle counter.

### 3.3. Methodology of the experimental study

*Overall laboratory experiment procedures*

Altogether, the lab study includes eight $CO_2$ corefloods in five cores. The sequence of procedures for each core is as follows (Fig. 5):

1. Undamaged gas permeability was measured by injecting $CO_2$ from gas cylinder (20). The flow rate of $CO_2$ was controlled at $51.02 \pm 0.02$ sccm (standard cubic centimetres per minute) by mass flow controller (17). The system was kept at back pressure of 300 psi by back pressure regulator (19). This was followed by core saturation in the coreholder by back-flow of 0.6 M NaCl solution at 0.02 mL/min.

2. Undamaged liquid permeability was measured by injecting 0.6 M NaCl solution (11) into the core at 2 mL/min using HPLC pump (10) until stabilization of pressure drop. Pressure drop across the core was measured by differential pressure transmitters (23-26).

3. Core drying was carried out by injecting $CO_2$ into the core at $50.86 \pm 0.12$ sccm at back pressure of 300 psi. Effluent particle concentration during brine and gas production, water saturation during drying process, and damaged gas permeability were measured. After the core drying process, core was re-saturated in the coreholder with by back-flow of 0.6 M NaCl solution at 0.02 mL/min.

4. Damaged liquid permeability was measured by injecting 0.6M NaCl solution into the core until stabilization of pressure drop across the core.

*Determination of brine saturation* During the process of core drying, the core was taken out from the coreholder every 24 hours, and its mass was measured (Fig. 5, photo 5) to calculate core saturation (photo 6). When the core mass has reached the initial dry mass, the core is completely dry.

*Collection of effluent particle concentration during $CO_2$ injection* Effluent particles were collected during three stages of core drying process: brine production, brine-gas production, and gas production.

During *brine production and brine-gas production*, the effluent samples were manually collected by visualising the level of liquid effluent collected in the sampling tubes (photo 7). Since the amount of effluent in each tube was insufficient for particle concentration (photo 9) and particle size distribution (PSD) measurement using the particle counter/sizer (photo 8), they were diluted by adding 10 mL of MilliQ deionized water. During *gas production*, the outlet tube was connected to a sealed Teflon collection bottle (photo 10) which allowed simultaneous continuous $CO_2$ flow and collection of produced fines (photo 11). Every 24 hours, the collection bottle was carefully rinsed with 10 mL MilliQ deionised water to collect produced fines during gas production and measure effluent particle concentration (photo 13) and PSD.

During the tests, we measured pressure drop across the core, breakthrough particle concentrations, and average water saturation, which are presented in the next section.

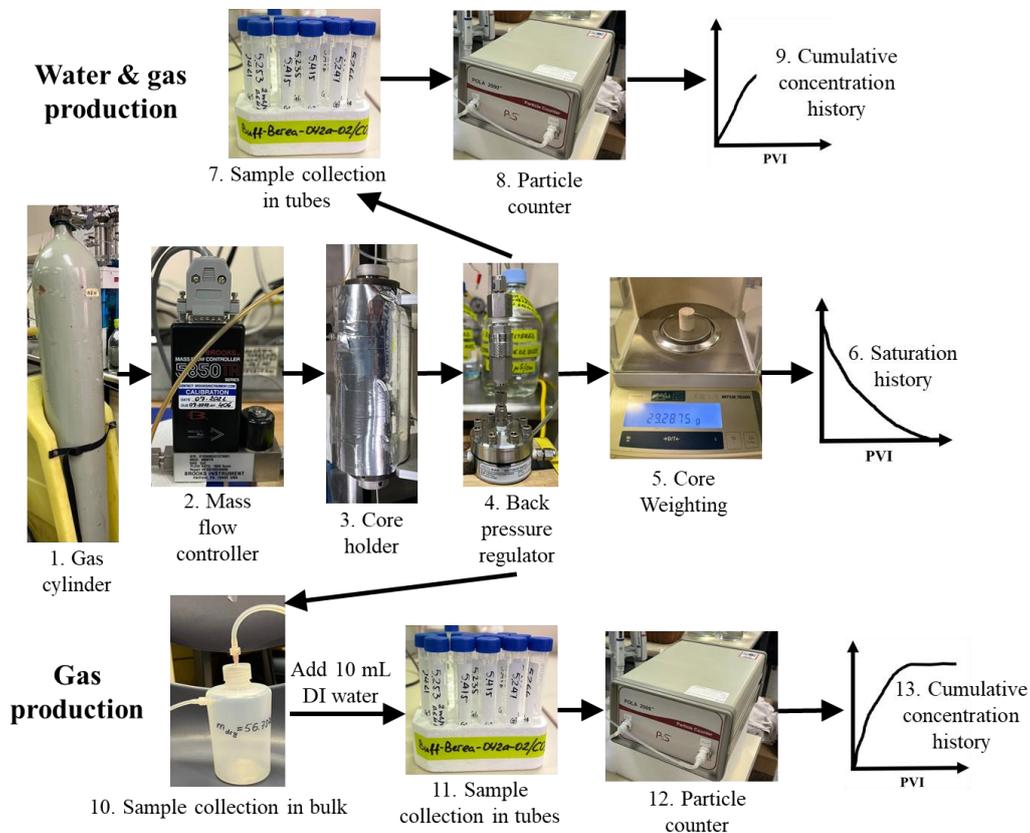

Figure 5. Photographic procedure of sample collection and measurements during $CO_2$ injection.

## 3.4. Experimental results

The results of four $CO_2$-flood tests in Berea 02 1st cycle, Berea 02 2nd cycle, Buff Berea 01 2nd cycle, and Bentheimer are presented in Figs. 6. Figs. S3, S4, S5, and S6 present coreflood data for Buff Berea 01 1st Cycle, Buff Berea 01 3rd cycle, Berea 01, and Buff Berea 02, respectively. Here left figures present histories of pressure drop across the cores (black curves), water saturation (blue curves), and cumulative concentration of produced fines (red curves); the time interval lasts from the beginning of injection to the moments of complete core drying. Right figures show the time zooms of those plots from the beginning of injection until 200-2000 PVIs, all – withing the water-gas production periods.

Table 1 presents the timing of main events in PVIs, $t_n$, and corresponding average saturations $S_w(t_n)$, where the low index corresponds to the event. Those comprise the period of water production $t_w$, n=w (second column), period of intensive fines production $t_i$, n=i (Fourth column), time of abrupt permeability decrease $t_k$, n=k (sixth column), the time where slow fines production starts $t_s$, n=s (eighth column), the overall drying period $t_f$, n=f, $S_w(t_f)=0$ (tenth column), and the ratio between final and initial pressure drops across the core (eleventh column). Figs. 7 and S7-S12 present particle size distributions in the effluent that correspond to those corefloods. Table 2 shows the ratios between the initial and post-mortem gas permeability for gas and for water (second and third columns, respectively).

Table 1. Times of the main events during all corefloods and corresponding water saturations

| Cores | Period of water production $t_w$, PVI | $S_w(t_w)$ | Intensive fines production $t_i$, PVI | $S_w(t_i)$ | Time of abrupt permeability increase $t_k$, PVI | $S_w(t_k)$ | Start of slow fines production $t_s$, PVI | $S_w(t_s)$ | Overall drying time $t_f$, PVI | $\Delta p/\Delta p_i$ |
|---|---|---|---|---|---|---|---|---|---|---|
| Berea 02 – 1st cycle | $6.5\times10^3$ | 0.65 | 272 | 0.997 | $0.94\times10^5$ | 0.47 | $1.3\times10^5$ | 0.45 | $3.7\times10^5$ | 0.43 |
| Berea 02 – 2nd cycle | $6.1\times10^3$ | 0.67 | 151 | 0.995 | $0.95\times10^5$ | 0.54 | $1.6\times10^5$ | 0.43 | $2.8\times10^5$ | 0.27 |
| Buff Berea 01 – 1st cycle | $2.7\times10^3$ | 0.62 | 146 | 0.998 | $0.21\times10^5$ | 0.62 | - | - | $1.9\times10^5$ | 1.11 |
| Buff Berea 01 – 2nd cycle | $1.7\times10^3$ | 0.68 | 170 | 0.995 | $0.24\times10^5$ | 0.69 | - | - | $2.1\times10^5$ | 1.10 |
| Buff Berea 01 – 3rd cycle | $2.9\times10^3$ | - | 120 | - | $0.71\times10^5$ | - | - | - | - | - |
| Bentheimer 01 | $1.9\times10^3$ | 0.68 | 120 | 0.998 | $0.27\times10^5$ | 0.68 | $0.54\times10^5$ | 0.33 | $1.4\times10^5$ | 2.21 |
| Berea 01 | - | 0.68 | - | - | $0.57\times10^5$ | 0.47 | - | - | $4.7\times10^5$ | 0.06 |
| Buff Berea 02 | $1.8\times10^3$ | - | 183 | - | $0.36\times10^5$ | - | $2.5\times10^5$ | - | $2.5\times10^5$ | - |

Table 2. Initial and damaged core permeabilities *vs* gas and water

| Core | Damaged gas permeability ratio ($k_{0,\ CO2}/k_{CO2}$) | Damaged liquid permeability ratio ($k_{0,\ brine}/k_{brine}$) |
|---|---|---|
| Berea 02 – 1st cycle | 4.75 | 2.65 |
| Berea 02 – 2nd cycle | 1.18 | 1.13 |
| Buff Berea 01 – 1st cycle | 1.67 | 1.28 |
| Buff Berea 01 – 2nd cycle | 5.0 | 1.06 |
| Buff Berea 01 – 3rd cycle | 1.0 | 1.0 |
| Bentheimer 01 | 2.07 | 0.97 |
| Berea 01 | 2.27 | 1.18 |
| Buff Berea 02 | 5.38 | 1.14 |

### 4. Laboratory data analysis and interpretation

Based on Figs. 6-9 and S3-S6, the following distinguished features of the histories of saturation (i, ii), pressure drop (iii, iv), and breakthrough concentration (v) have been observed.

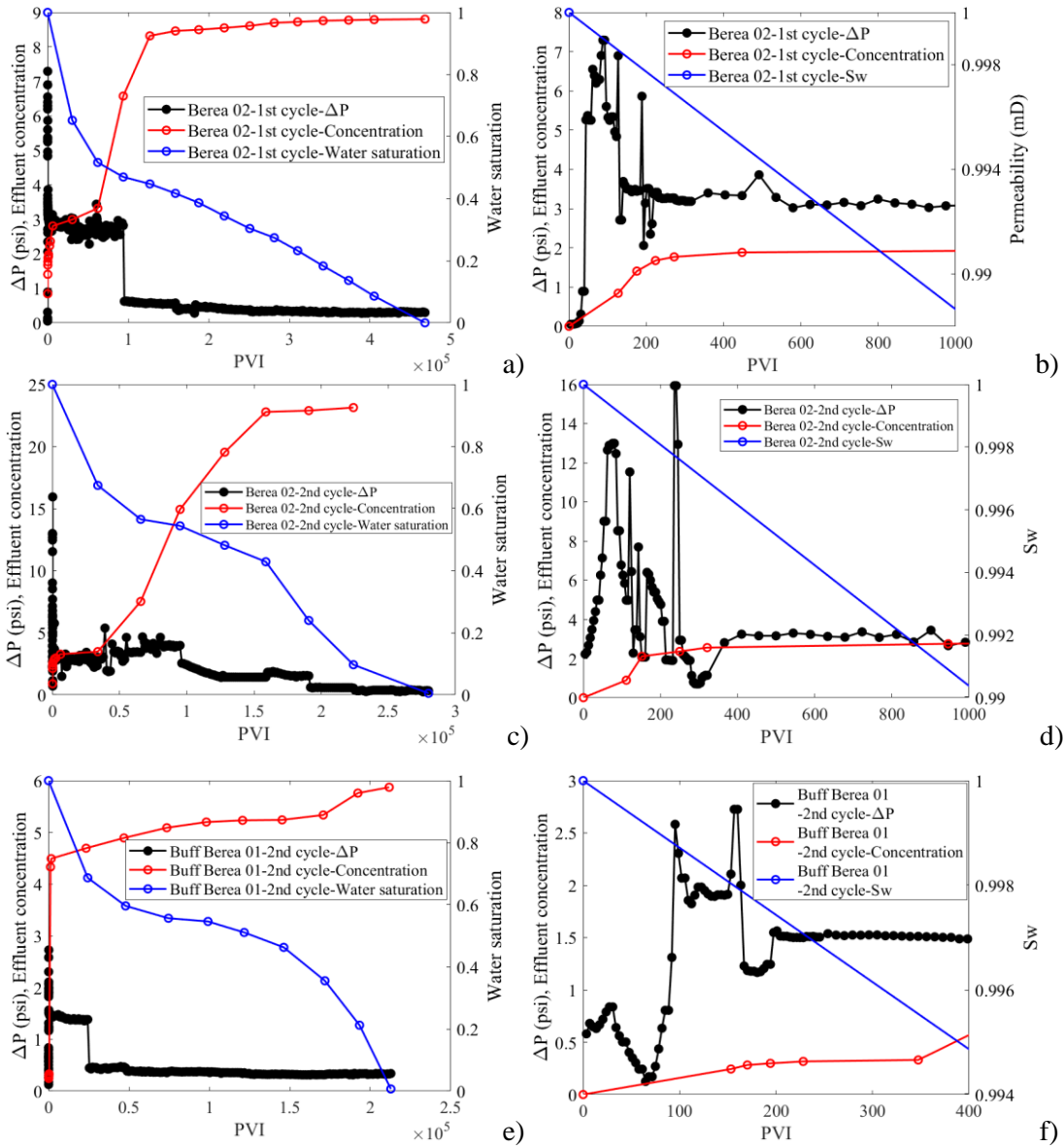

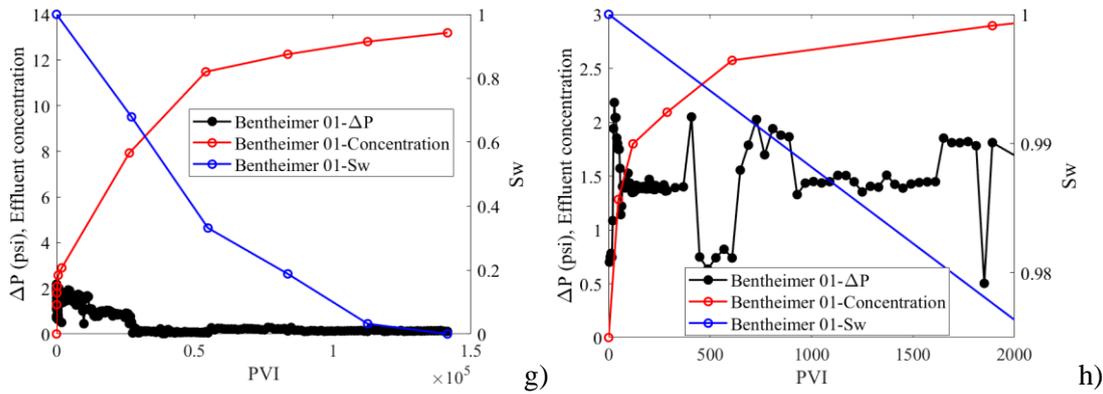

Figure 6. The results on pressure drop, cumulative effluent concentration, and water saturation: (a, b) 1st cycle of Berea 02; (c, d) 2nd cycle of Berea 02; (e, f) 2nd cycle of Buff Berea 01; (g,h) Bentheimer; (a, c, e, g) during the overall coreflood; (b, d, e, h) at the initial stage.

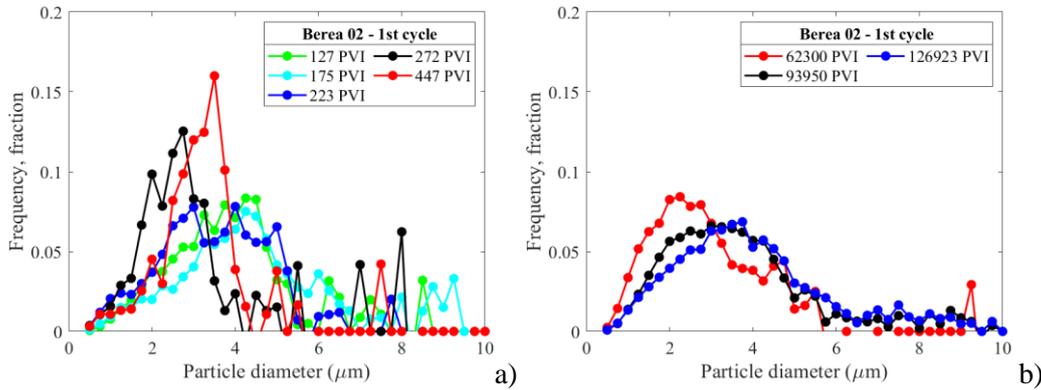

Figure 7. Effluent particle size distribution during (a) gas and water production; (b) gas production only for the first cycle of drying of Berea 02 core.

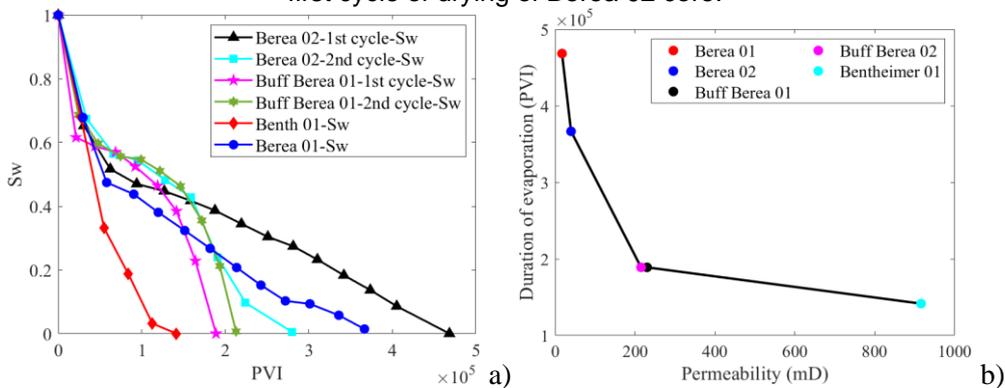

Figure 8. Dynamics of core drying: a) Brine saturation as a function of PVI during drying of four sandstone cores (Berea 02, Buff Berea 01, Bentheimer 01, Berea 01); b) Duration of evaporation vs permeability.

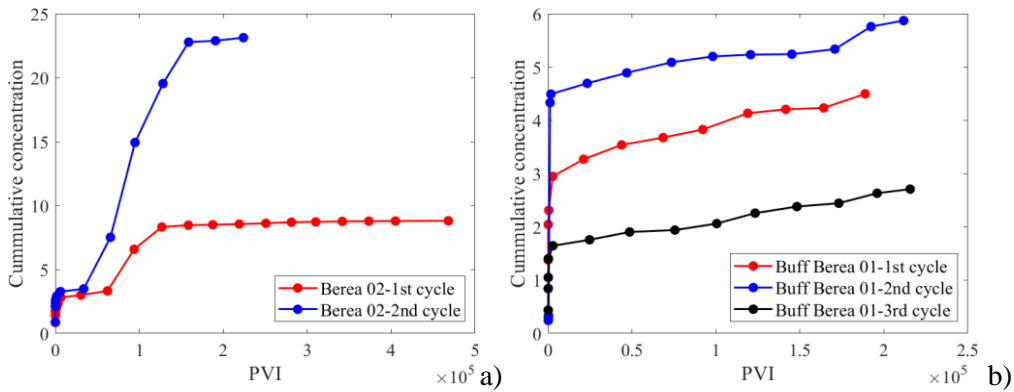

Figure 9. The comparison of cumulative concentrations of different cycles: (a) Berea 02; b) Buff Berea 01.

(i) *Water saturation history* Water is produced during 1700-3000 of PVIs of gas; saturation at the end of water production varies withing the narrow interval 0.62-0.68 (Table 1). The evaporation rate, which is the slope of saturation history, start declining at the moment of $0.2\times10^5$-$0.5\times10^5$ PVI. Sometimes, it increases at the end of evaporation (Figs. 6a, 6c, and S3).

Extremely high connate saturation is explained by gas viscosity ($\mu_g=1.5\times10^{-5}$ Pa.s) which is significantly lower than water viscosity ($\mu_w=10^{-3}$ Pa.s), yielding unfavourable mobility ratio at the gas-water front and consequent viscous fingering. Even long after gas breakthrough, the microscale sweep coefficient is low, and the gas channels form a bundle of parallel paths connecting core inlet and outlet [27].

The rock is water-wet, so during two-phase flow, gas-water menisci move towards small pores (Fig. 1a). So, during single-phase flow with evaporation, the menisci radii decrease (Fig. 1b). At both stages, the total menisci area decreases, yielding the decrease in evaporation rate. However, saturation decrease during continuous gas injection yields the increase of the number of viscous fingers, resulting in the increase in water-gas interface. Water evaporation into each finger from cross-flow channels and dead-end pores also yields some increase in water-gas interface. So, non-monotonic evaporation rate is explained by two competitive effects, which are in odds with each other: evaporation in pores in order of decreasing menisci size and interface, and also increase in the total area of the gas-water interface [28, 29]. At the beginning, the effect of radii decrease dominates, while at the end the evaporation rate is controlled mostly by the increase in the gas-water interface.

(ii) *Finite evaporation period* The water saturation gradually decreases during evaporation and reaches zero in a finite time, i.e., there is no asymptotical drying where saturation tends to zero as time tends to infinity (Fig. 8a). Finite or infinite evaporation period is defined by whether the power $n$ in the rate of evaporation or chemical reaction exceeds one or not [30, 31]. The overall evaporation time has order of magnitude $0.1$-$0.5\times10^6$ PVI, which agrees with other laboratory studies [32, 33].

Fig. 8a shows the tendency of higher evaporation rate in highly permeable cores: saturation history for highly-permeable Bentheimer is represented by the lowest red curve, while the black, blue and light blue curves for low permeability Berea cores are above the green and purple curves for medium-permeability Buff Berea cores during the bulk of test periods. Table 1 and Fig. 8b show the final-evaporation time: the higher is the permeability the faster occurs the full evaporation. This effect is attributed to high irregularity of low-permeable rock surface, the presence of multiple crevices and asperities, housing small water patches with small menisci. Evaporation from the joints particle – flat substrate, which is likely to be in large pores of highly permeable cores, happens faster.

(iii) *Pressure drop behaviour*   The pressure drop at intermediate time scale either decrease (Figs. 6a, 6e, 6g, S3a, S4a, S6a) or increase (Fig. 7c). The same is observed at the beginning of water-production periods (all right Figs. 6, S4, and S5). However, at large times, from the moment of the pressure drop jump down and until full evaporation, the pressure drop monotonically decreases. During the overall test periods, the pressure drop across the core can either increase or decrease (Table 2). The ratio between the finite and initial pressure drops varies from 0.97 to 5.38.

Let us explain the non-monotonic pressure-drop history. From the beginning of injection and up to the moment of abrupt pressure drop fall, the pressure drop increases due to mobilisation, migration, and size exclusion of natural reservoir fines [1, 2, 4]. Simultaneously, water saturation decrease during gas injection yields increase in gas permeability and pressure drop decrease. So, the pressure drop behaviour during short and intermediate times is explained by two competitive factors which are in odds to each other: gas saturation increase and fines migration. Pressure drop increase corresponds to fines-migration and straining domination, while the saturation decrease domination yields the decrease in the pressure drop. Due to the unavailability of $k_{rg}(s)$ during unstable displacement, it is not possible to distinguish between the pressure drop increase by formation damage and decrease due to an increasing $k_{rg}(s)$. Late monotonic pressure drop decrease until its stabilisation coincides with vanishing of particles at the effluent (the last stage of fines production), which can be observed during all corefloods (all left Figs. 6 and S3-S6). So, slow monotonic pressure drop decline at large times is attributed to slow water evaporation and increase in gas network conductivity.

(iv) *Abrupt pressure-drop decline*   At the moment $0.2 \times 10^5$-$1.0 \times 10^5$ PVI, pressure drop abruptly decreases. Yet sometimes, 2-3 abrupt pressure drop falls are observed (Fig. 6c, S3a and S5a).

Consider the set of pores saturated by non-wetting phase in an infinite pore network during stable displacement of wetting water. The non-wetting phase is immobile for saturation above interstitial value $S_{wi}$, where the displacing phase forms a set of finite clusters. Upon reaching the threshold value, the non-wetting phase conductivity grows as $(S_{wi}-s)^n$, $n<1$ [34-36]. Unstable displacement corresponds to percolation in finite network, i.e. to invasion percolation. Separate invading fingers provide some conductivity for the non-wetting displacing phase; saturation decrease below the threshold value $S_{wi}$ yields significant conductivity growth.

Water saturation during gas injection decreases due to gradual displacement of water in conductive pore paths and evaporation of water from the transversal-to-gas-flow or dead-end pores, into the conductive channels. The evaporation results in expanding of side branches of gas channels, so-called beard in the percolation theory. We attribute abrupt pressure drop decrease in Figs. 6a, 6e, 6g, S4, and S6 by creation of conductive gas network at some threshold saturation.

(v) *Breakthrough fines concentration* Left Figs. 6, S3-S6 show three typical periods of the cumulative particle breakthrough curves: sharp increase at the beginning of injection during 150-500 PVI that corresponds to high fines production; slower increase until $0.4 \times 10^5$-$0.7 \times 10^5$ PVI corresponding to moderate fines production rates, and, finally, slow vanishing of the particles in the produced fluid. Besides, a plateau was observed for low-permeable Berea 02 for both cycles. The plateau disappears for the cores with higher permeability.

From the beginning of injection during the initial stage of the water-production period, a large number of particles is produced at the outlet (Figs. 6, S3-S6). Those fines are detached by capillary forces exerted on the particles by the gas-water menisci that travel through the pores during two-phase displacement (Fig. 1a). The plots of the ratio between the capillary and maximum electrostatic forces in Fig. 3a are calculated for conditions of these coreflood tests. The plots show that the capillary force is significantly higher than the attaching electrostatic DLVO force, which explains high fines-production rate during two-phase flow period.

During evaporation, particles are held to the surface by water pendular rings (Fig. 1b); here the capillary force is an attaching force. The particle sizes in the tests vary from 1 μm to 3 μm. Fig. 3b shows the ratio between detaching and attaching torques. The details of calculation are presented in section S1. The detaching forces are drag and lift; the attaching forces are capillary, electrostatic, and gravity. The lever arm is equal to the radius of the contact particle-substrate deformation circle, which is calculated using Hertz's theory [37]; the load is equal to the total of all normal forces. A particle detaches if the detaching torque exceeds the attaching torque. The blue curve in Fig. 3b corresponds to flow velocity in our tests, indicating that the particles remain attached for all positions and volumes of the brine pendular rinds. Increasing the velocity to 0.1 m/s results in the torque ratio almost reaching unity. Only velocity increase up to 0.5 m/s results in fines detachment at small filing angles, i.e., where the rock is almost dry. Therefore, the capillary attraction by the water pendular rings is significantly higher than the detaching drag.

Given sufficient time, the pendular rings completely evaporate, resulting in detachment due to the drag force exerted by the flowing gas. Fig. 3b shows that the particles can be detached by the gas drag for velocities that exceed the velocities in these tests by one or two orders of magnitude. Finally, the fines cannot be detached during the second and third stages of $CO_2$ flooding (Figs. 1b and 1c, respectively).

However, unexpectedly, a significant fraction of the fines was produced when water production does not occur anymore. The fines production rate during water production is one-two orders of magnitude higher than that during the dry production period, but water production period is also one-two orders of magnitude lower the dry production period. So, the cumulative fines concentrations produced during the first and later stages have the same order of magnitude. Let us explain the phenomenon. During unstable gas displacement, the fingers fill in the porous space in order of decreasing of the path permeability. Moreover, in swept areas, non-wetting gas fills the pores in order of decrease of their radii in accordance with the increase of capillary pressure threshold resisting gas invasion under the pressure gradient. So, different stages shown in Figs. 1a, 1b, and 1c occur simultaneously in different pores, i.e., the stages overlap in the overall core.

Gas velocity in large pores can significantly exceed its average value, yielding the increase in detaching drag, while two-phase displacement and evaporation are going on in other core patterns.

Now consider thin pores filled by water after the end water production, so those pores do not form an infinite cluster. Consider two schematics of water evaporation from those pores into an infinite conducting gas cluster: (i) water-gas menisci moves inside the water-filled pores; (ii) water-gas level in those pores decreases. In the case (i), the attached particles are removed from the pore surfaces; being removed into an unstable position on the top of asperity or of the rough surface, the particles are moved by drag inside gas phase.

Berea rocks are mixed-wet; $CO_2$ injection increases the contact angle. Fig. 3a shows that detaching capillary force at high values of the water filling angle highly exceeds the maximum of attaching DLVO, i.e. the fines can be detached at high water levels of the interface in the pore. This explains fines detachment in the case (ii) when water production has already stopped.

Table 2 shows that the damage ratio between the initial and stabilised permeabilities (i.e., before and after each test) is higher in gas than in water. This is attributed to dissolution of the *precipitated salt* during rock re-saturation by water after 1st and 2nd cycles. Some precipitated salt is observed in the core inlet and outlet (Fig. 10). Thus, permeability damage for gas is explained by both salt precipitation and fines migration, while for water the permeability decrease occurs only by strained fines.

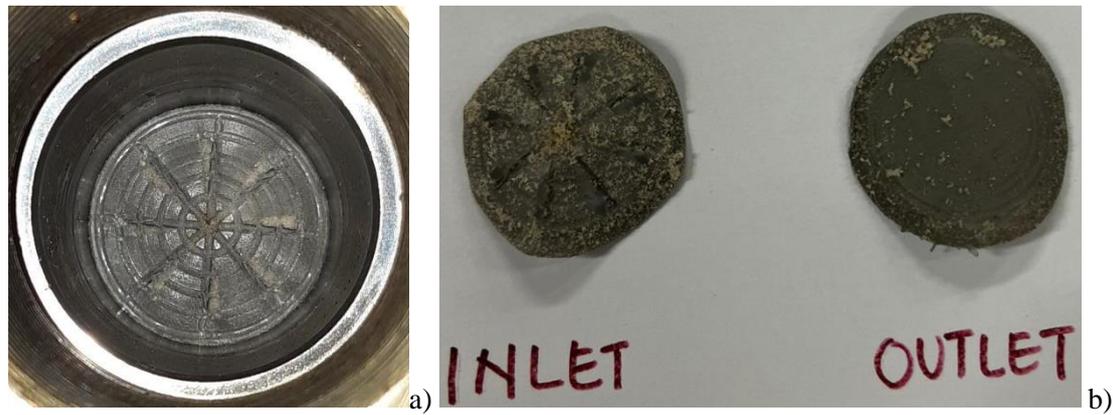

Figure 10. Photo of (a) salt precipitation at inlet flow distributor and mesh; (b) salt precipitation on inlet and outlet mesh.

Figs. 6c, 6d, 6e, 6f, and S4 show that significant *fines production* occurs *during secondary and tertiary $CO_2$ floods*, while the primary floods have been carried out until full water evaporation, where the attaching capillary force disappears, and no detachable particles are left on the rock surface. This effect is attributed to different geometry of gas-filled porous space at various test stages. During low-viscosity gas injection and unstable displacement, stochastic viscous fingering occurs, i.e. flow paths do not repeat under an identical flow conditions. Besides, fines migration yields the permeability damage of the gas paths; in the next gas flood after the core re-saturation by water, gas enters more permeable paths, yielding the changing flow paths. Figs. 9 compare fines production during different flood cycles. Later floods can yield lower or higher fines production, no dependency has been observed. The above speculations are applicable for both authigenic [38] and detrital [39] fines.

*Freshwater flood* after several high-salinity brine injections yields significant permeability decline due to fines production (Fig. S14). This is explained by dissolution of thin pendular rings of residual brine, captured by capillary force at the particle - rock surface junctions – in the injected fresh water. Fig. 2 shows that DLVO attraction in high-salinity environment is significantly stronger than that for fresh water. The particles that remain attached after the first and second flushes by highly saline brine and $CO_2$ become detached by drag against the weakened DLVO attraction.

The effects of *particle sizes* on their detachment and size exclusion have been intensively discussed [40, 41]. Preferable mobilisation of large particles occurs for detrital fines, which are detached by drag against electrostatic force, and for authigenic fines that are detached by breakage [38, 39]. The same result is expected during increase of gas saturation, where fines are gradually detached by capillary force [22]. However, no correlation has been observed in Fig. S13, which presents time variation of mean particle diameter during all 8 corefloods. Probably, more complex models rather than reflecting detachment of spheroid particles from plane substrates alone must be developed to interpret the data.

5. Discussions

*Particle size distribution*   The detrital fines are detached against the attaching DLVO forces during flow with velocity increase in order of their size decreasing, i.e., first detach large particles, then smaller and smaller [40, 41]. The same detachment order has been observed for authigenic fines, which are detached by breakage [38]. Therefore, shift of PDFs of the effluent fines to the left while the injection rate increases, should be expected to occur in Figs. 7 and S7-S12, which is not the case. Fig. S13 also does not show mean particle size decrease during flow velocity increase.

Moreover, large particles are selectively captured by straining, yielding timely particle size decrease at the effluent, which is not observed. Permeability decreases during $CO_2$ floods, presented in Table 2, suggests the important role of straining in the capture of the detached and migrating fines. However, fines attachment may occur in the order of particle size increase, i.e., first small particles are attached. The quantitative analysis of these three factors, which compete and are in odds with each other, may explain absence of tendency in BTCs behaviour.

*Modelling*   The qualitative explanations of the physics phenomena provided in Section 4 must be confirmed by the mathematical modelling, matching the lab data, and tuning the model parameters. Development of the governing equations to capture the effects described in Section 4 will require significant theoretical efforts.

Unstable displacement of water by gas will be described by invasion percolation for calculating gas saturation, and by either effective medium or critical-path theory for calculating conductivities (phase permeability) for gas and water, like it was performed for stable displacement in works [36, 42, 43]. Kinetics of evaporation is described by the relaxation equation where the rate is proportional to the difference between the equilibrium and current vapour concentrations in gas [28, 29].

Population balance model, integrated with two-phase flow equations, will reflect the role of different size particles and pores [44, 45]. Here both attachment and straining will be considered. The modelling will match PDFs for effluent particles from Figs. 7 and S7-S12. The exact upscaling will yield the mass balance equations for averaged concentrations of the particles and pores [44, 45], coupled with the equation for saturation, allowing matching BTCs along with the histories of mean saturations and pressure drops (Figs. 6 and S3-S6).

PDFs for produced fines can be alternatively treated by random walk models [46, 47], continuous upscaling [48, 49] or Boltzmann's equation [50].

## 6. Conclusions

The set of lab tests on brine displacement by $CO_2$ in sandstone cores and the data interpretation by DLVO theory and capillary phenomena allows drawing the following conclusions.

Torque/force balance on attached particles show that the movable particles can be detached by passing menisci but remain attached by the pendant rings of the residual brine and during gas flow in dry pores. All stages are present simultaneously in the core due to unstable displacement, micro heterogeneity, and distributed pore sizes.

Large period of gas-water production – thousands of PVIs - and high saturation when it ends up is explained by extremely unfavourable gas-water mobility ratio yielding unstable displacement of water by gas and intensive viscous fingering.

Non-monotonic variation of the evaporation rate observed in all floods is explained by two competitive factors during saturation decrease: decrease in the menisci radii in water-wet rocks yielding decrease of the was-water interface; increase of the number of fingers and evaporation into the fingers resulting in the interface increase.

Finite evaporation period of the order of magnitude $10^5$ PVIs is observed at all floods. The higher is the permeability, the higher the evaporation rate, and the faster is the rock drying.

Non-monotonic pressure-drop behaviour observed in almost all floods is explained by increasing gas permeability due to saturation decrease, from one side, and by the gas permeability decrease due to fines lifting with the following straining. The abrupt pressure-drop decrease in almost all tests is explained by reaching the percolation threshold by gas saturation where gas forms an infinite connected cluster of the conducting pore paths.

The sharp decrease of fines production during the water-gas production period and its slow decrease during evaporation is explained by fines detachment during two-phase flow and their attachment by capillary and electrostatic forces by the residual water bridges and in dry pores. Yet, fines production occurs continuously throughout the gas production period.

Fines production during secondary and tertiary floods of the same cores is explained by damaging the gas-flow paths in previous floods and establishing news flow paths in the following floods.

Significant fines production and permeability decline during post mortem fresh water flooding occurs due to brine bridge dissolution in fresh water, where the particle-rock attraction is significantly weaker than in brine; the particle-rock attraction is fresh water is also significantly weaker than in gas environment. This explains why the fines retaining after evaporation stages given by Figs. 1b and 1c can be removed by the slow freshwater flux.

The ratio between the initial and final permeabilities vs CO$_2$ is higher than that vs brine because brine dissolves the precipitated salt.

**Declaration of competing interest**

The authors declare that they have no known competing financial interests or personal relationships that could have appeared to influence the work reported in this paper.

**References**

[1] Khilar, K.C. and Fogler, H.S., 1998. *Migrations of fines in porous media (Vol. 12)*. Springer Science & Business Media.

[2] Civan, F., 2023. *Reservoir formation damage: fundamentals, modeling, assessment, and mitigation*. Gulf Professional.

[3] Johnson, W. P and Pazmino, E. F. 2022. Colloid (Nano- and Micro-Particle) Transport and Surface Interaction in Groundwater. *The Groundwater Project*. 111 pp. ISBN: 978-1-77470-070-9. DOI: https://doi.org/10.21083/978-1-77470.

[4] Yang, Y., Yuan, W., Hou, J. and You, Z., 2022. Review on physical and chemical factors affecting fines migration in porous media. *Water Research*, 214, p.118172.

[5] Chequer, L., Bedrikovetsky, P., Badalyan, A. and Gitis, V., 2020. Brine level and mobilisation of colloids in porous media. *Advances in Water Resources*, 143, p.103670.

[6] Qi, M., Li, Y., Moghanloo, R.G., Guo, T. and Wu, N., 2023. Applying deep bed filtration theory to study long-term impairment of fracture conductivity caused by reservoir fines. *Geoenergy Science and Engineering*, 231, p.212253.

[7] Iglauer, S., Pentland, C.H. and Busch, A., 2015. CO2 wettability of seal and reservoir rocks and the implications for carbon geo-sequestration. *Water Resources Research*, *51*(1), pp.729-774.

[8] Baban, A., Keshavarz, A., Amin, R. and Iglauer, S., 2022. Impact of Wettability Alteration on CO2 Residual Trapping in Oil-Wet Sandstone at Reservoir Conditions Using Nuclear Magnetic Resonance. *Energy & Fuels*, *36*(22), p.13722-13731.

[9] Alanazi, A., Baban, A., Ali, M., Keshavarz, A., Iglauer, S. and Hoteit, H., 2023. Residual trapping of CO2, N2, and a CO2-N2 mixture in Indiana limestone using robust NMR coreflooding: Implications for CO2 storage. *Fuel*, *353*, p.129221.

[10] Saraf, S. and Bera, A., 2021. A review on pore-scale modeling and CT scan technique to characterize the trapped carbon dioxide in impermeable reservoir rocks during sequestration. *Renewable and Sustainable Energy Reviews*, *144*, p.110986.

[11] Gholami, R. and Raza, A., 2022. CO2 sequestration in sandstone reservoirs: How does reactive flow alter trapping mechanisms? *Fuel*, *324*, p.124781.

[12] Ge, J., Zhang, X., Liu, J., Almutairi, A. and Le-Hussain, F., 2022. Influence of capillary pressure boundary conditions and hysteresis on CO2-water relative permeability. *Fuel*, *321*, p.124132.

[13] Zareei, D., Rostami, B. and Kostarelos, K., 2022. Petrophysical changes of carbonate rock related to CO2 injection and sequestration. *International Journal of Greenhouse Gas Control*, *117*, p.103648.

[14] Zareei, D., Rostami, B. and Kostarelos, K., 2022. Petrophysical changes of carbonate rock related to CO2 injection and sequestration. *International Journal of Greenhouse Gas Control*, *117*, p.103648.

[15] Sokama-Neuyam, Y.A., Yusof, M.A. and Owusu, S.K., 2022. CO2 Injectivity in Deep Saline Formations: The Impact of Salt Precipitation and Fines Mobilization. *Carbon Sequestration*.

[16] Yusof, M.A.M., Neuyam, Y.A.S., Ibrahim, M.A., Saaid, I.M., Idris, A.K. and Mohamed, M.A., 2022. Experimental study of CO2 injectivity impairment in sandstone due to salt precipitation and fines migration. *Journal of Petroleum Exploration and Production Technology*, *12*(8), pp.2191-2202.

[17] Ge, J., Zhang, X., Othman, F., Wang, Y., Roshan, H. and Le-Hussain, F., 2020. Effect of fines migration and mineral reactions on CO2-water drainage relative permeability. *International Journal of Greenhouse Gas Control*, *103*, p.103184.

[18] Ge, J., Zhang, X. and Le-Hussain, F., 2022. Fines migration and mineral reactions as a mechanism for CO2 residual trapping during CO2 sequestration. *Energy*, *239*, p.122233.

[19] Sazali, Y.A., Sazali, W.M.L., Ibrahim, J.M., Graham, G. and Gödeke, S., 2020. Investigation of fines migration for a high-pressure, high-temperature carbonate gas reservoir offshore Malaysia. *Journal of Petroleum Exploration and Production Technology*, *10*, pp. 2387-2399.

[20] Elimelech, M., Gregory, J., Jia, X., Williams, R.A., 1995. *Particle Deposition & Aggregation*. Butterworth-Heinemann.

[21] Israelachvili, J.N., 2022. Surface forces. In *The Handbook of Surface Imaging Visualization* (pp. 793-816). CRC Press.

[22] Nguyen, C., Loi, G., Russell, T., Shafian, S.M., Zulkifli, N.N., Chee, S.C., Razali, N. and Zeinijahromi, A., 2022. Well inflow performance under fines migration during water-cut increase. *Fuel*, *327*, p.124887. [23] Xie, Q., Saeedi, A., Delle Piane, C., Esteban, L. and Brady, P.V., 2017. Fines migration during CO2 injection: Experimental results interpreted using surface forces. *International Journal of Greenhouse Gas Control*, *65*, pp.32-39.

[24] Ali, M., Awan, F.U.R., Ali, M., Al-Yaseri, A., Arif, M., Sánchez-Román, M., Keshavarz, A. and Iglauer, S., 2021. Effect of humic acid on CO2-wettability in sandstone formation. *Journal of colloid and interface science*, 588, pp.315-325.

[25] Arif, M., Abu-Khamsin, S.A. and Iglauer, S., 2019. Wettability of rock/CO2/brine and rock/oil/CO2-enriched-brine systems: Critical parametric analysis and future outlook. *Advances in colloid and interface science*, 268, pp.91-113.

[26] Yuan, B. and Moghanloo, R.G., 2019. Analytical modeling nanoparticles-fines reactive transport in porous media saturated with mobile immiscible fluids. *AIChE Journal*, 65(10), p.e16702.

[27] Lake, L.W., Johns, R., Rossen, B. and Pope, G.A., 2014. *Fundamentals of enhanced oil recovery (Vol. 1, p. 1)*. Richardson, TX: Society of Petroleum Engineers.

[28] Yiotis, A.G., Boudouvis, A.G., Stubos, A.K., Tsimpanogiannis, I.N. and Yortsos, Y.C., 2004. Effect of liquid films on the drying of porous media. *AIChE Journal*, *50*(11), pp.2721-2737.

[29] Yortsos, Y.C. and Stubos, A.K., 2001. Phase change in porous media. *Current opinion in colloid & interface science*, *6*(3), pp.208-216.

[30] Altree-Williams, A., Brugger, J., Pring, A., 2019. Coupled reactive flow and dissolution with changing reactive surface and porosity. *Chemical Engineering Science*, 206, pp.289-304.

[31]. Nazaikinskii, V.E., Bedrikovetsky, P.G., Kuzmina, L.I. and Osipov, Y.V., 2020. Exact Solution for Deep Bed Filtration with Finite Blocking Time. *SIAM Journal on Applied Mathematics*, 80(5), pp.2120-2143.

[32] Peysson, Y., 2012. Permeability alteration induced by drying of brines in porous media. *The European Physical Journal-Applied Physics*, *60*(2), p.24206.

[33] Peysson, Y., Fleury, M. and Blázquez-Pascual, V. (2011) Drying rate measurements in convection-and diffusion-driven conditions on a shaly sandstone using nuclear magnetic resonance. *Transport in Porous Media* 90, 1001-1016.

[34] Shklovskiĭ, B.I. and Éfros, A.L.V., 1975. Percolation theory and conductivity of strongly inhomogeneous media. *Soviet Physics Uspekhi*, *18*(11), p.845.

[35] Hunt, A., Ewing, R. and Ghanbarian, B., 2014. *Percolation theory for flow in porous media* (Vol. 880). Springer.

[36] Bedrikovetsky, P., 2013. Mathematical theory of oil and gas recovery: with applications to ex-USSR oil and gas fields (Vol. 4). *Springer Science & Business Media*.

[37] Timoshenko, S.P. and Goodier, J.N., *Theory of elasticity* 3rd edition 1970.

[38] Hashemi, A., Borazjani, S., Nguyen, C., Loi, G., Khazali, N., Badalyan, A., Yang, Y., Dang-Le, B. and Russell, T., 2023. Particle detachment in reservoir flows by breakage due to induced stresses and drag. *International Journal of Rock Mechanics and Mining Sciences*, *172*, p.105591.

[39] Hashemi, A., Nguyen, C., Loi, G., Khazali, N., Yang, Y., Dang-Le, B., Russell, T. and Bedrikovetsky, P., 2023a. Colloidal detachment in porous media: Stochastic model and upscaling. *Chemical Engineering Journal*, 474, p.145436.

[40] Yang, Y., Siqueira, F.D., Vaz, A., Badalyan, A., You, Z., Zeinijahromi, A., and Carageorgos, T., 2018. Fines migration in aquifers and oilfields: laboratory and mathematical modelling. *Flow and Transport in Subsurface Environment*, pp.3-67.

[41] Russell, T., Chequer, L., Borazjani, S., You, Z., and Zeinijahromi, A., 2018. Formation damage by fines migration: Mathematical and laboratory modeling, field cases. In *Formation Damage During Improved Oil Recovery* (pp. 69-175). Gulf Professional Publishing.

[42] Salimi, H., Bruining, J. and Joekar-Niasar, V., 2020. Comparison of modified effective-medium approximation to pore-network theory for relative permeabilities. *Journal of Petroleum Science and Engineering*, *184*, p.106594.

[43] Naik, S., Gerke, K.M. and You, Z., 2021. Application of percolation, critical-path, and effective-medium theories for calculation of two-phase relative permeability. *Physical Review E*, *103*(4), p.043306.

[44] Bedrikovetsky, P., 2008. Upscaling of stochastic micro model for suspension transport in porous media. *Transport in porous media*, *75*, pp.335-369.

[45] Bedrikovetsky, P., Osipov, Y., Kuzmina, L. and Malgaresi, G., 2019. Exact upscaling for transport of size-distributed colloids. *Water Resources Research*, 55(2), pp.1011-1039.

[46] Shapiro, A.A., 2007. Elliptic equation for random walks. Application to transport in microporous media. *Physica A: Statistical Mechanics and its Applications*, *375*(1), pp.81-96.

[47] Yuan, H. and Shapiro, A.A., 2010. Modeling non-Fickian transport and hyperexponential deposition for deep bed filtration. *Chemical Engineering Journal*, *162*(3), pp.974-988.

[48] Shapiro, A.A., 2021. Continuous upscaling and averaging. *Chemical Engineering Science*, *234*, p.116454.

[49] Shapiro, A.A., 2022. Continuous upscaling of the 3D diffusion equation in a heterogeneous medium. *Chemical Engineering Science*, *248*, p.117247.

[50] Shapiro, A.A. and Wesselingh, J.A., 2008. Gas transport in tight porous media: Gas kinetic approach. *Chemical Engineering Journal*, *142*(1), pp.14-22.

# SUPPLEMENTARY MATERIAL

## S1. Forces exerting the attached particle

*Electrostatic DLVO force*

The electrostatic force is expressed via the energy potential $V$:

$$F_e = -\frac{dV}{dh} \tag{1}$$

which is the total of three components: van der Waals, electric double layer, and Born repulsion [1, 2]:

$$V = -\frac{A_{132} r_s}{6h} + 64\pi\varepsilon_0 \varepsilon_r r_s \left(\frac{k_B T}{z_v e}\right)^2 \gamma_p \gamma_s \exp(-\kappa h) + \frac{A_{132} r_s \sigma_c^6}{1260 h^7} \tag{2}$$

Here $r_s$ is the particle radius (varying from 0.92 μm to 2.89 μm), $h$ is the distance between the bottom of the particle and the substrate, $\varepsilon_o$ is the permittivity of free space (8.85×10$^{-12}$ C$^2$N$^{-1}$m$^{-2}$), $\varepsilon_r$ is the relative permittivity of water (78.46), $k_B$ is the Boltzmann constant (1.38×10$^{-23}$ J·K$^{-1}$), $T$ is the absolute temperature, $z_v$ is the ion valency (1 for NaCl), $e$ is the charge of an electron (1.6×10$^{-19}$ C), $\sigma_c$ is the atomic collision coefficient (0.5 nm). The inverse Debye length, $\kappa$, is [2]:

$$\kappa = 0.73 \times 10^8 \sqrt{\sum C_{mi} z_i^2} \tag{3}$$

where $C_{mi}$ is the molar concentration of the $i$-th ion in moles/m$^3$, $z_i$ is the valence of the $i$-th ion. The reduced zeta potentials for the particle, $\gamma_p$, and substrate, $\gamma_s$, are given by:

$$\gamma_p = \tanh\left(\frac{ze\zeta_p}{4k_B T}\right); \gamma_s = \tanh\left(\frac{ze\zeta_s}{4k_B T}\right) \tag{4}$$

The zeta potential for the kaolinite particle, $\zeta_p$, is equal to -23.4 mV in a 0.6M NaCl solution and -52.5 mV in DI water, and the zeta potential for sandstone, $\zeta_s$, is equal to -18.1mV in a 0.6M NaCl solution and -62.2 mV in DI water. The expression for the Hamaker constant, $A_{132}$, is given as [1]:

$$A_{132} = \frac{3}{4} k_B T \left(\frac{\varepsilon_1 - \varepsilon_3}{\varepsilon_1 + \varepsilon_3}\right)\left(\frac{\varepsilon_2 - \varepsilon_3}{\varepsilon_2 + \varepsilon_3}\right) + \frac{3h_c v_e}{8\sqrt{2}} \frac{(\eta_1^2 - \eta_3^2)(\eta_2^2 - \eta_3^2)}{(\eta_1^2 + \eta_3^2)^{1/2}(\eta_2^2 + \eta_3^2)\left\{(\eta_1^2 + \eta_3^2)^{1/2} + (\eta_2^2 + \eta_3^2)\right\}} \tag{5}$$

where numbered subscripts in the formula represent the particle (1), substrate (2), and solution (3). The static dielectric constants for the particle and substrate are taken for kaolinite ($\varepsilon_1$ =11.18) and sandstone ($\varepsilon_2$ =3.76) [3, 4]. The value for the fluid, $\varepsilon_3$, is equal to 78.46 for water and 1.01 for $CO_2$ [5]. The refractive indices of the kaolinite particle and sandstone are given as $\eta_1$=1.56 and $\eta_2$=1.55 [3, 6]. The refractive index for water is $\eta_3$=1.33, and for $CO_2$ is $\eta_3$=1.00 [5]. The absorption frequency, $v_e$, is equal to $3.0\times10^{15}s^{-1}$, and the Planck constant, $h_c$, is $6.62\times10^{-34}$ J·s. The calculated $A_{132}$ is equal to $1.57\times10^{-20}$ J for the kaolinite-sandstone interaction in a water environment and $8.56\times10^{-20}$ J in a $CO_2$ environment.

*Capillary force acting on a spherical particle*

When a moving interface passes by the spherical particle (during drainage or imbibition), the capillary force acting on the particle induced by the interface is given as [7]:

$$F_c = -2\pi r_s \sigma \sin\omega \sin(\omega+\theta) \tag{6}$$

where $\sigma$ is the $CO_2$-water interfacial tension (0.058 Nm$^{-1}$ at 300 psi) [8], $\theta$ is the particle contact angle (60.3° for kaolinite particle) [9], and $\omega$ is the filling angle (see Fig.1a).

The capillary force exerting on the particle by the liquid pendular ring situated between the particle and the substrate is (see Fig.1b) [10]:

$$F_c = 4\pi\sigma r_s \cos(\theta) - \frac{4\pi\sigma h r_s \cos(\theta)}{\sqrt{h^2 + V_r/(\pi r_s)}} - 2\pi r_s \sigma \sin\omega \sin(\omega+\theta) \tag{7}$$

where $h$ is the distance between the particle and substrate, which we set as 0.2 nm (distance between particle and substrate when particle is attached in the primary minimum) for all subsequent calculations, and $V_r$ is the volume of the liquid pendular ring between the particle and substrate. The relationship between the filling angle, $\omega$, and $V_r$ is expressed as [10]:

$$V_r = \frac{\pi r_s^3 \omega^2}{4} + \pi h \omega r_s^2 \tag{8}$$

*Formulas for Hydrodynamic force acting on a spherical particle*

Expressions for drag force and lift force acting on a spherical particle attached to a plane substrate during laminar flow are given as [11, 12]:

$$F_d = 1.7009 \times 6\pi\mu r_s U \tag{9}$$

$$F_l = 81.2\sqrt{\rho\mu}(r_s U)^{1.5} \tag{10}$$

where $\mu$ is the fluid viscosity (10$^{-3}$ Pa.s for water and $1.5\times10^{-5}$ Pa.s for $CO_2$) [13], $U$ is the velocity acting on the particle centre, and $\rho$ is the fluid density (10$^3$ kg/m$^3$ for water and 39.43 kg/m$^3$ for $CO_2$ at 300 psi) [14].

*Hertz theory for elastic particle deformation*

When the net vertical force is pointing downwards, the particle will be deformed, creating a deformation area between the particle and substrate. The radius of the deformed area is calculated using Hertz theory [15, 16]. In our case, the deformed area radius is equal to the lever arm for the normal forces, $l_n$:

$$l_n = \sqrt[3]{\frac{3F_n r_s}{4E_Y^*}} \tag{11}$$

$$\frac{1}{E_Y^*} = \frac{1-v_p^2}{E_{Yp}} + \frac{1-v_g^2}{E_{Yg}} \tag{12}$$

where $F_n$ is the normal force acting on the particle, which is the sum of the electrostatic force, capillary force, and lift force in this scenario, $E_Y^*$ is the effective Young's modulus, $E_{Yp}$ (1.64×10$^{10}$ Pa) and $E_{Yg}$ (2.42×10$^{10}$ Pa) represent the Young's moduli of the kaolinite particle and sandstone, respectively, $v_p$ (0.30) is the Poisson's ratio of the particle, and $v_g$ (0.20) is the Poisson's ratio of the sandstone [17-19].

**S2. Core properties**

Table S1. Core dimensions and properties

| Sandstone core | L, cm | D, cm | $\phi_{imb}$ | $k_{0,brine}$, mD | $k_{0,CO2}$, mD |
|---|---|---|---|---|---|
| Berea 01 | 2.548 | 2.529 | 18.09 | 17.10 | 50 |
| Berea 02 | 2.554 | 2.533 | 18.22 | 39.73 | 95 |
| Buff Berea 01 | 2.368 | 2.498 | 21.84 | 230 | 250 |
| Buff Berea 02 | 2.368 | 2.498 | 21.84 | 215 | 211 |
| Bentheimer 01 | 2.384 | 2.517 | 22.17 | 916 | 580 |

*Mineralogical rock composition* is determined from XRD analysis shown below in Fig. S1. The composition is presented in Table S2.

Table S2. Mineral core compositions

| Cores | Mineralogical composition, % (w/w) | | | | | |
|---|---|---|---|---|---|---|
| | Quartz | Kaolinite | Microcline | Muscovite | Albite | Calcite |
| Berea 01 | 95.67 | 1.39 | N/A | 1.72 | 0.26 | 0.95 |
| Berea 02 | | | | | | |
| Buff Berea 01 | 93.51 | 2.02 | 1.87 | 1.12 | 1.47 | N/A |
| Buff Berea 02 | | | | | | |
| Bentheimer 01 | 93.60 | 0.05 | 2.48 | 3.86 | N/A | N/A |

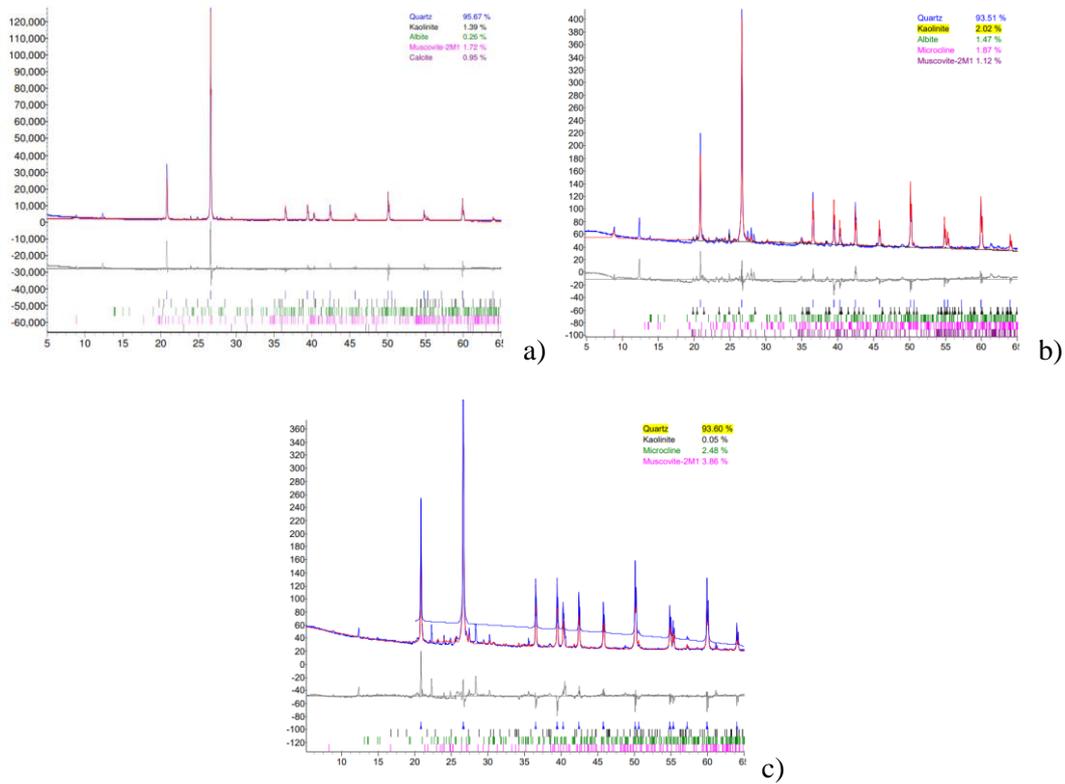

Figure S1. XRD analysis of the cores used in the study: a) Berea, b) Bluff Berea, c) Bentheimer

**S3. Laboratory set-up**

Fig. 4 shows the schematic of the experimental; Fig. S2 shows its photo. Sandstone core *(1)* is placed inside an elastic Viton sleeve *(2)*. Two stainless steel fluid distributors *(3)* are used to fix the core in place by compression. A vertical high-pressure coreholder *(4)* (model RCH, CoreLab, USA) accommodates the Viton sleeve and sandstone core. This setup is designed to hold up to 5000 psi maximum pressure (model 87-6-5, High Pressure Equipment Company, USA). An overburden pressure generator *(5)* develops an overburden pressure by compressing distilled brine *(6)* to prevent leakage between the outer surface of the core and the Viton sleeve. The overburden pressure is measured by an absolute pressure transmitter *(7)* (model PA-33X, KELLER AG fur Druck-messtechnik, SWITZERLAND). Inlet and outlet pressures in the core-holder are measured by pressure transmitter *(8)* and *(9)* (model PA-33X, KELLER AG fur Druckmesstechnik, SWITZERLAND). A Prep-36 high-performance liquid chromatography (HPLC) pump *(10)* pumps brine solutions *(11)* through the sandstone core via manual valves *(12)* and *(13)*. Gaseous $CO_2$ from a compressed gas cylinder *(14)* via a gas pressure regulator *(15)*, a manual valve *(16)* and a mass flow controller *(17)* (model 5850TR, BROOKS) is injected into a sandstone core via manual valves *(13 and 18)*. A back-pressure regulator *(19)* (EQUILIBAR) maintains the pressure inside the core at 300 psi by using compressed air *(20)* via gas pressure regulator *(21)* and a manual valve *(22)*. The pressure drop across the sandstone core is measured by four differential pressure transmitters *(23-26)* (model EJX 110A, Yokogawa Electric Corporation, JAPAN) with four different measuring ranges (0 – 1, 0 – 14.5, 0 – 72.5 and 0 – 2350 psi). Switching between differential pressure transmitters is carried out via four manual valves *(27-30)*. Electrolytic conductivity sensor *(31)* (Microelectrodes, model 8-900) sends signal to signal transmitter *(32)* (eDAQ Conductivity isoPod) to record electrolytic conductivity data. An ADAM-4019+ inlet data acquisition module *(33)* (ADVANTECH™, TAIWAN) and RS-232/RS/485 signal conditioner ADAM-5060 *(34)* (ADVENTECH™) receive signals from all transmitters in real time and feed them into a PC-based data acquisition system *(35)*. A custom-built data acquisition software (ADVANTECH ADAMView Ver. 4.25 application builder) records all experimental parameters in real-time mode and performs all necessary calculations. Effluent suspensions via a manual valve *(36)* are collected in plastic sampling tubes *(37)* located in the sampling carousel *(38)*. Concentration of effluent samples collected in effluent samples are measured by a POLA-2000 particle counter/sizer *(39)* (Particle and Surface Sciences, Australia).

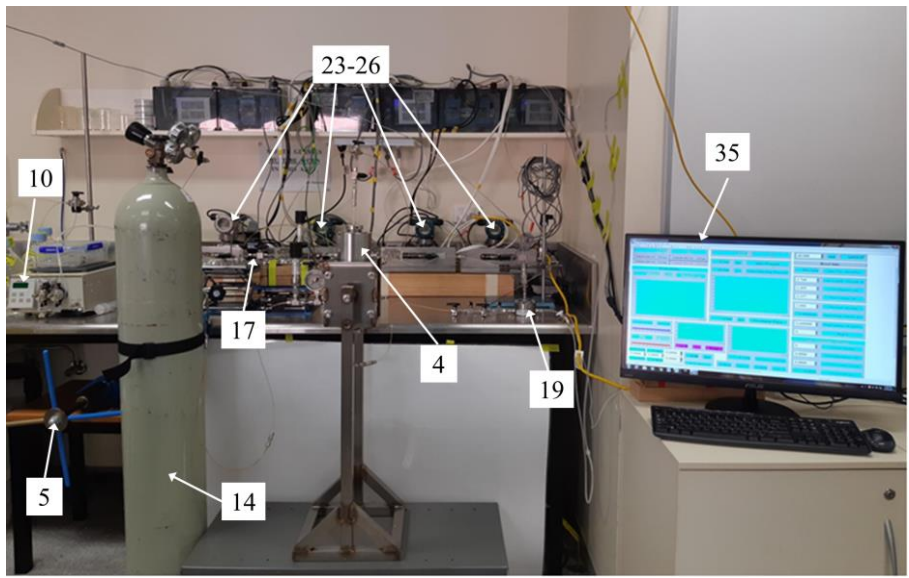

Figure S2. Photo of the experimental setup.

## S4. Coreflooding results

Here we present the measurements for four corefloods: Buff Berea 01 – 1st cycle, Buff Berea 01 – 3rd cycle; Berea 01, and Buff Berea 02, presented in Figs. S3, S4, S5, and S6, respectively. Fig. 6 in the main text shows the data from other four floods: 1st cycle of Berea 02, 2nd cycle of Berea 02, 2nd cycle of Buff Berea 01, and Bentheimer. Then follow the data on produced particle size distribution (Figs. S7-S12).

**Buff Berea 01 – 1st cycle**

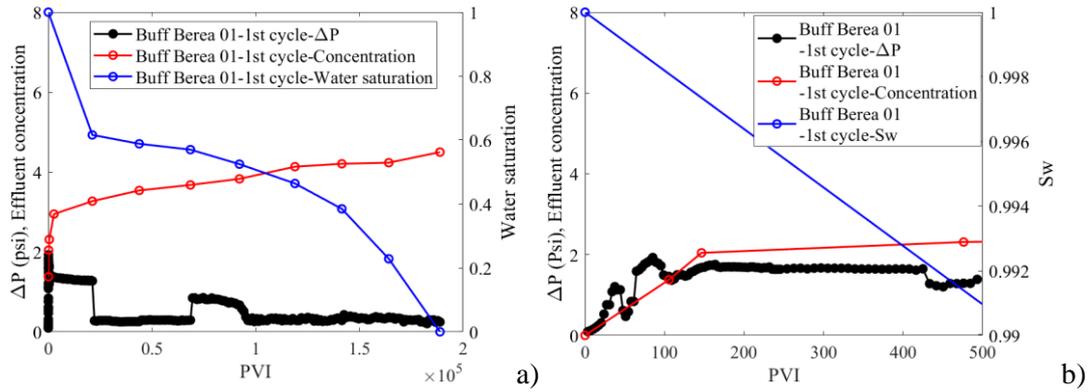

Figure S3. The results for variation of pressure drop, cumulative effluent concentration and water saturation (a) across the core; (b) at the initial stage during the first cycle of drying of Buff Berea 01 core.

**Buff Berea 01 – 3rd cycle**

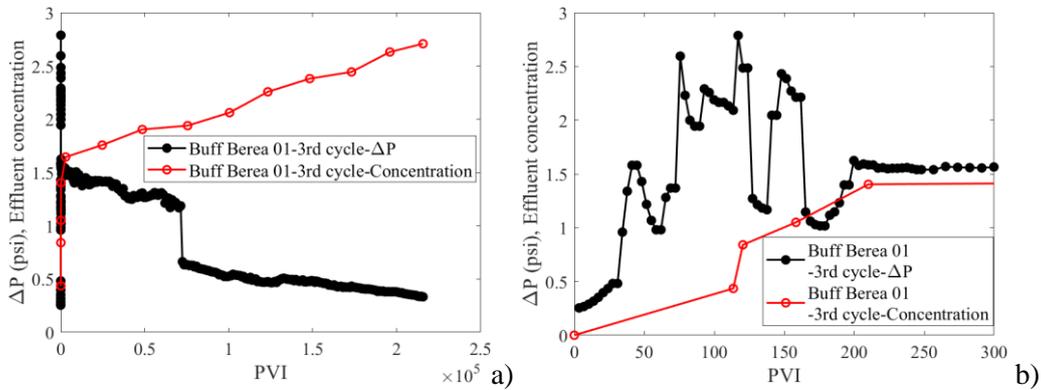

Figure S4. The results for variation of pressure drop and cumulative effluent concentration (a) across the core; (b) at the initial stage during the third cycle of drying of Buff Berea-01 core.

**Berea 01**

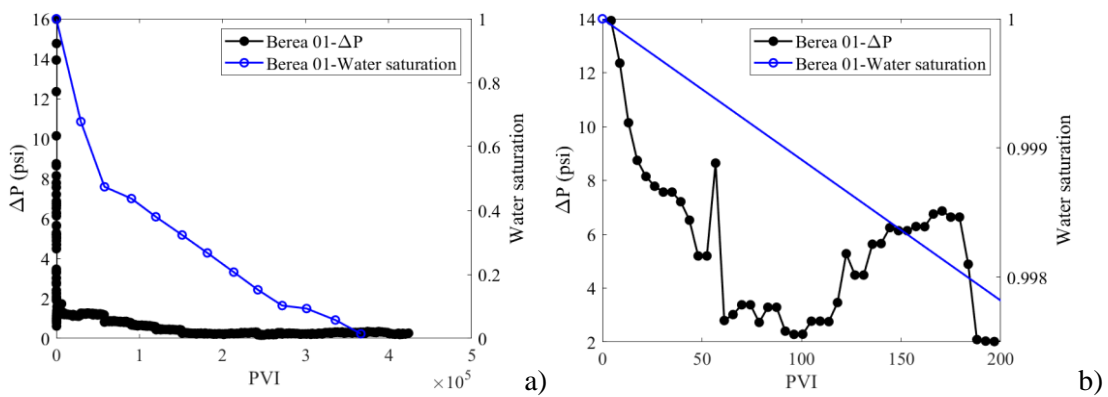

Figure S5. The results for variation of pressure drop and gas permeability (a) across the core; (b) at the initial stage during drying of Berea 01 core.

**Buff Berea 02**

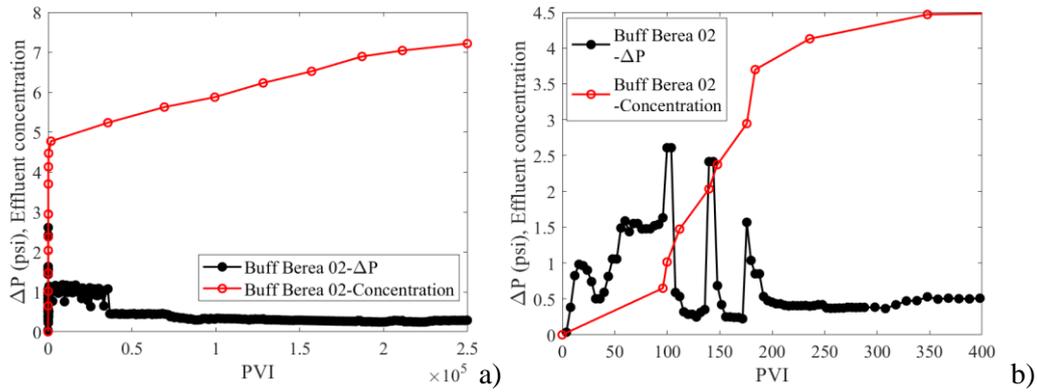

Figure S6. The results for variation of pressure drop and cumulative effluent concentration (a) across the core; (b) at the initial stage during the third cycle of drying of Buff Berea 02 core.

*Particle size distributions during corefloods* is an important information on the micro-scale parameters, which can be extracted by matching the data by population balance models. For the first cycles of injections in core Berea 02, the particle size distributions are presented in Figs. 7a and 7b. In this section, we present PDFs for particles for second cycle of Berea 02 (Fig S7), first coreflood cycle of Buff Berea 01 (Fig. S8), second cycle of Buff Berea 01 (Fig. S9), third cycle of Buff Berea 01 (Fig. S10), Bentheimer (Fig. S11), and Buff Berea 02 (Fig. S12). Figs. a) present PDFs during water-gas production, b) – afterward during gas production.

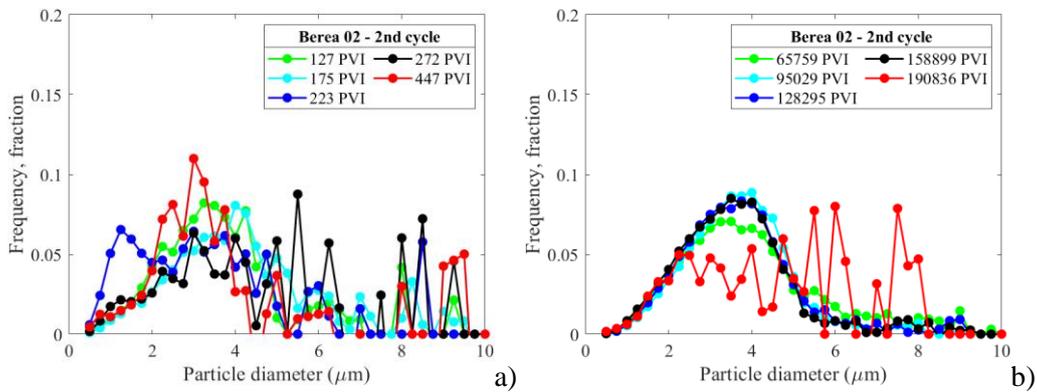

Figure S7. Effluent particle size distribution during (a) gas and water production; (b) gas production only for the second cycle of drying of Berea 02 core.

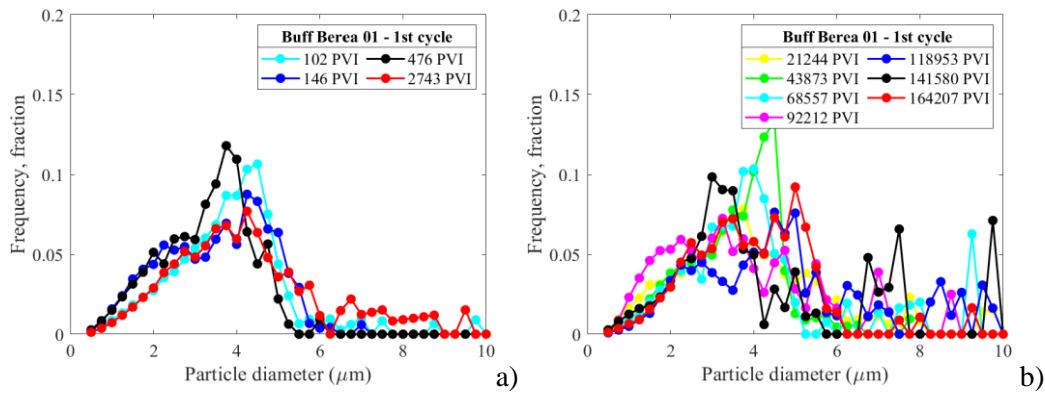

Figure S8. Effluent particle size distribution during: (a) gas and water production; (b) gas production only for the first cycle of drying of Buff Berea 01 core.

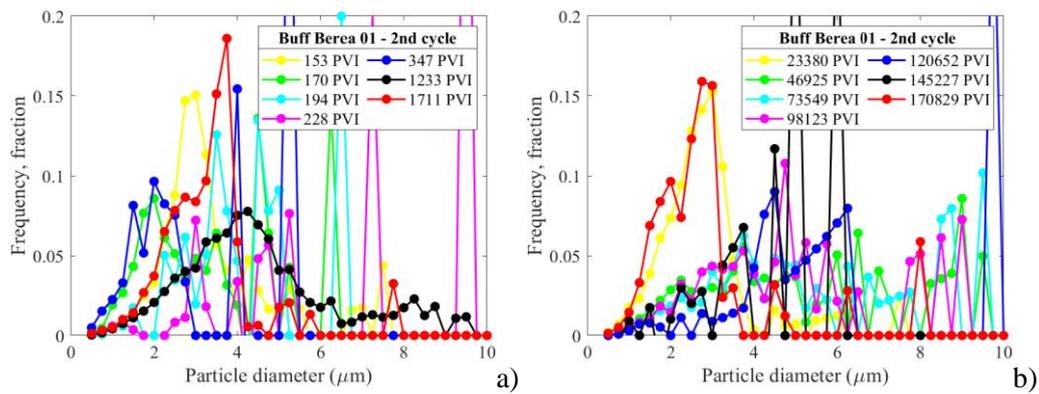

Figure S9. Effluent particle size distribution during: (a) gas and water production; (b) gas production only for the second cycle of drying of Buff Berea 01 core.

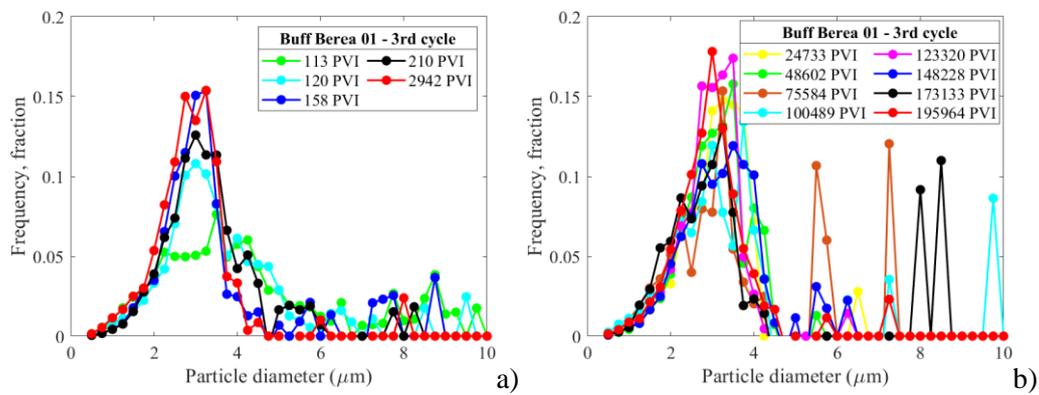

Figure S10. Effluent particle size distribution during: (a) gas and water production; (b) gas production only for the third cycle of drying of Buff Berea 01 core.

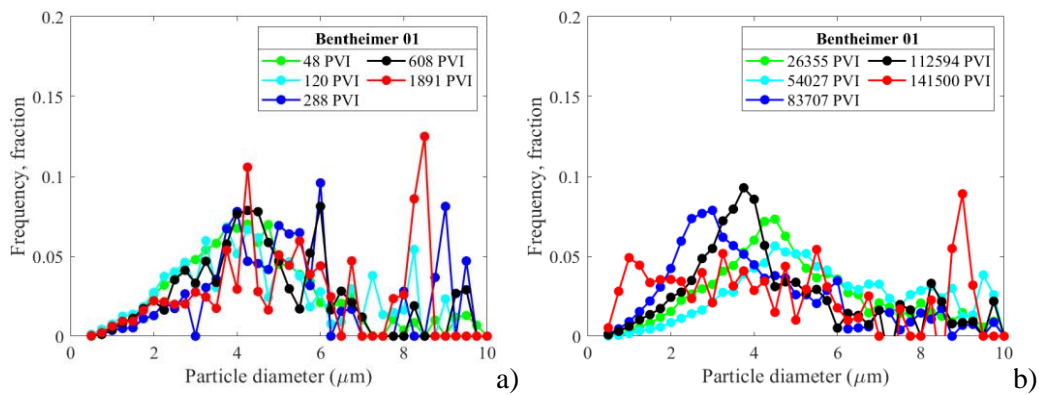

Figure S11. Effluent particle size distribution during: (a) gas and water production; (b) gas production only during the drying of Bentheimer 01 core.

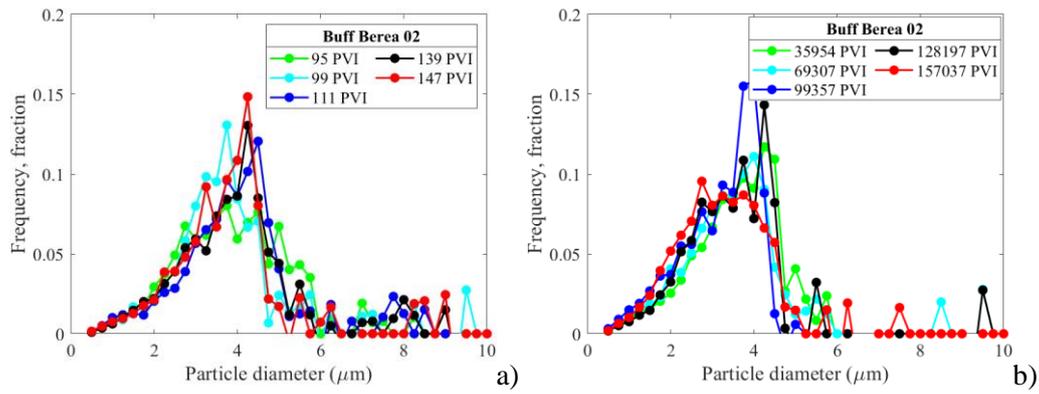

Figure S12. Effluent particle size distribution during: (a) gas and water production; (b) gas production only during the drying of Buff Berea 02 core.

No timely tendency in size PDFs of the produced fines was observed.

Fig. S13 presents time variation of the mean particle diameter with time for all five corefloods, where also no pattern of behaviour was observed.

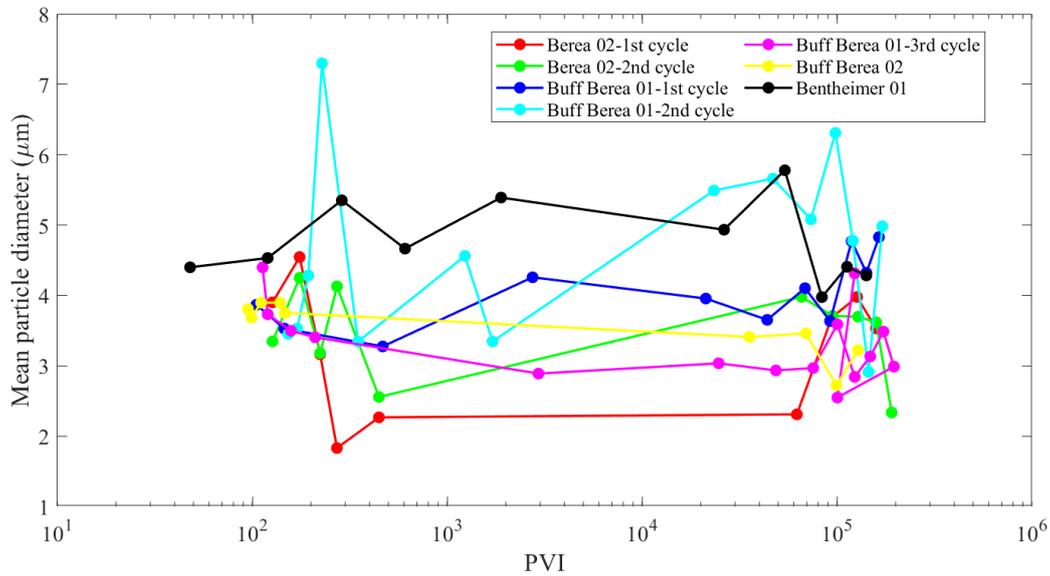

Figure S13. Variation of mean particle diameter for all cores.

Fig. S14 shows drastic permeability decline during fresh water injection after second cycle flooding of core Berea 02, which is attributed to massive fines detachment and migration.

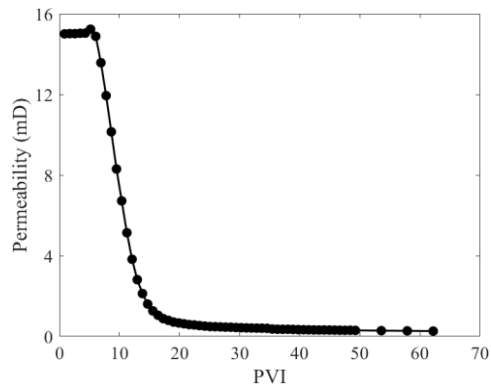

Figure S14. Permeability decline during deionized water injection after the second cycle of core drying of Berea 02 core.

**References**

[1] Hubbard, A.T., 2022. *The Handbook of surface imaging and visualization*. CRC press.

[2] Elimelech, M., Gregory, J. and Jia, X., 2013. *Particle deposition and aggregation: measurement, modelling and simulation*. Butterworth-Heinemann.

[3] Khilar, K.C. and Fogler, H.S., 1998. *Migrations of fines in porous media* (Vol. 12). Springer Science & Business Media.

[4] Rosenholtz, J.L. and Smith, D.T., 1936. The dielectric constant of mineral powders. *American Mineralogist: Journal of Earth and Planetary Materials*, *21*(2), pp.115-120.

[5] Kohmun, K., Limsuwan, P., Ranusawud, M. and Buranasiri, P., 2020. Effect of the carbon dioxide concentration on the Refractive Index of air in a long gauge block interferometer. *Instrumentation Science & Technology*, *48*(4), pp.350-360.

[6] Weidler, P.G. and Friedrich, F., 2007. Determination of the refractive index of particles in the clay and sub-micrometer size range. *American Mineralogist*, *92*(7), pp.1130-1132.

[7] Sharma, P., Flury, M. and Zhou, J., 2008. Detachment of colloids from a solid surface by a moving air–water interface. *Journal of colloid and interface science*, *326*(1), pp.143-150.

[8] Espinoza, D.N. and Santamarina, J.C., 2010. Water-CO2-mineral systems: Interfacial tension, contact angle, and diffusion—Implications to CO2 geological storage. *Water resources research*, *46*(7).

[9] Wang, J., Cao, Y., Xing, Y., Gui, X. and Li, G., 2022. Study on the wetting behavior between oil droplets and kaolinite substrate based on interaction force measurement and high-speed dynamic visualization. *Colloid and Interface Science Communications*, *46*, p.100585.

[10] Rabinovich, Y.I., Esayanur, M.S. and Moudgil, B.M., 2005. Capillary forces between two spheres with a fixed volume liquid bridge: theory and experiment. *Langmuir*, *21*(24), pp.10992-10997.

[11] O'neill, M.E., 1968. A sphere in contact with a plane wall in a slow linear shear flow. *Chemical Engineering Science*, *23*(11), pp.1293-1298.

[12] Saffman, P.G., 1965. The lift on a small sphere in a slow shear flow. *Journal of fluid mechanics*, *22*(2), pp.385-400.

[13] Fenghour, A., Wakeham, W.A. and Vesovic, V., 1998. The viscosity of carbon dioxide. *Journal of physical and chemical reference data*, *27*(1), pp.31-44.

[14] Duschek, W., Kleinrahm, R. and Wagner, W., 1990. Measurement and correlation of the (pressure, density, temperature) relation of carbon dioxide II. Saturated-liquid and saturated-vapour densities and the vapour pressure along the entire coexistence curve. *The Journal of Chemical Thermodynamics*, *22*(9), pp.841-864.

[15] Bradford, S.A., Torkzaban, S. and Wiegmann, A., 2011. Pore-scale simulations to determine the applied hydrodynamic torque and colloid immobilization. *Vadose Zone Journal*, *10*(1), pp.252-261.

[16] Derjaguin, B.V., Muller, V.M. and Toporov, Y.P., 1975. Effect of contact deformations on the adhesion of particles. *Journal of Colloid and interface science*, *53*(2), pp.314-326.

[17] Molina, O., Vilarrasa, V. and Zeidouni, M., 2017. Geologic carbon storage for shale gas recovery. *Energy Procedia*, *114*, pp.5748-5760.

[18] Wang, Z., Wang, H. and Cates, M.E., 2001. Effective elastic properties of solid clays. *Geophysics*, *66*(2), pp.428-440.

[19] Húlan, T. and Štubňa, I., 2020. Young's modulus of kaolinite-illite mixtures during firing. *Applied Clay Science*, *190*, p.105584.